\documentclass[aps,twocolumn,prl,floatfix,groupedaddress,nofootinbib,notitlepage,showpacs,superscriptaddress]{revtex4-2}
\usepackage{amsmath}
\usepackage{times,bm,bbm,bbold,graphicx,graphics,amssymb,amsmath,amsfonts,dsfont,hyperref,color}
\usepackage{mathrsfs,cancel}
\usepackage[utf8]{inputenc}
\usepackage{graphicx}
\usepackage{soul,xcolor}
\setstcolor{red}
\definecolor{mycolor}{rgb}{0.1, 0.1, 0.7}
\usepackage{dsfont}
\usepackage{url}
\usepackage[vlined,ruled]{algorithm2e}
\usepackage{natbib}
\DeclareFontFamily{OT1}{pzc}{}
\DeclareFontShape{OT1}{pzc}{m}{it}%
{<-> s * [1.25] pzcmi7t}{}
\hypersetup{
	plainpages=true,
	breaklinks=true,
	hypertexnames=false,
	pageanchor=true,
	colorlinks=true,
	linkcolor={mycolor},
	citecolor={mycolor},
	urlcolor={mycolor},
	anchorcolor={black}
}
\usepackage{verbatim}

\usepackage{appendix}
\usepackage{physics}
\begin{document}

\title{ Environment-Assisted Generation of Non-Gaussian Wavepacket Quantum States}
\author{Maryam Khanahmadi}
\email{m.khanahmadi@chalmers.se}
\affiliation{Department of Microtechnology and Nanoscience, Chalmers University of Technology, 412 96 Gothenburg, Sweden}
\author{Klaus M{\o}lmer}
\affiliation{Niels Bohr Institute, University of Copenhagen, Jagtvej 155A, DK-2200 Copenhagen, Denmark} 

\date{\today}
\begin{abstract}

Generating non-Gaussian states and converting them into traveling wavepackets is crucial yet challenging for scalable, fault-tolerant quantum computing. We present a hardware-efficient approach that simultaneously achieves both tasks by combining
an engineered nonlinear dissipation with a linear transmission loss from a superconducting circuit to a waveguide. This combination of dissipative channels leverages low-order
interactions to induce a high-order nonlinearity, enabling deterministic emission of a wide range of non-Gaussian, error-correctable states, such as Schr\"odinger cat states, GKP states, and pair-cat states. We identify experimental superconducting circuit platforms and realistic parameter regimes for our proposal. 
\end{abstract}

\maketitle

\textit{Introduction---}
Encoding quantum information in states that populate propagating wave packets is essential for communication in a quantum internet \cite{kimble2008quantum} and for coupling of remote quantum processors in scalable architectures for quantum computing \cite{PhysRevLett.120.200501,campagne2018deterministic}. Using photons or phonons as carriers, the losses in transmission lines, along with decoherence and dephasing errors in the emitters and receivers, limit the fidelity of the desired remote quantum operations. One possibility to overcome these effects is to encode the quantum information in the logical basis of error-correctable quantum states, permitting recovery of the information despite the possible errors \cite{axline2018demand,PhysRevApplied.12.044067,yang2024deterministic}. Different non-Gaussian bosonic states such as Schrödinger cat states, binomial states, and Gottesman-Kitaev-Preskill (GKP) grid states have been proposed to realize fault-tolerant quantum communication and computing \cite{PhysRevLett.111.120501,PhysRevLett.119.030502,terhal2020towards,PhysRevA.101.012316,joshi2021quantum,bourassa2021blueprint}.
Preparation of quantum information into the logical basis of the Schrödinger cat states has been extensively explored in the optical regimes, either by photon subtraction from squeezed states \cite{ourjoumtsev2006generating,wakui2007photon,PhysRevLett.121.143602} or by utilizing the interaction with single atoms \cite{deleglise2008reconstruction,hacker2019deterministic}. In the microwave regime, the Kerr-nonlinearity in Josephson Junctions has been used for deterministic generation of cat states in quantum resonator eigenmodes \cite{vlastakis2013deterministically,mirrahimi2014dynamically,mirrahimi2016cat,he2023fast}.

Previous works have explored architectures that include a tunable coupler to transfer the prepared stationary state to a waveguide 
\cite{pfaff2017controlled, campagne2018deterministic, axline2018demand, khanahmadi2023multimode}. Such couplers may introduce unwanted non-linear interactions, and the longer duration of the separate preparation and release processes reduces the output quantum state fidelity due to dissipation.  In this letter, we directly generate traveling non-Gaussian states by parametric driving of a non-linear resonator undergoing constant linear loss to a transmission waveguide. In this approach, the release is faster, and the profile of the parametric drive determines the shape of the propagating wave packet and eliminates the need for tunable couplers. 

\begin{figure}[ht!]
\includegraphics[width=.5\textwidth]{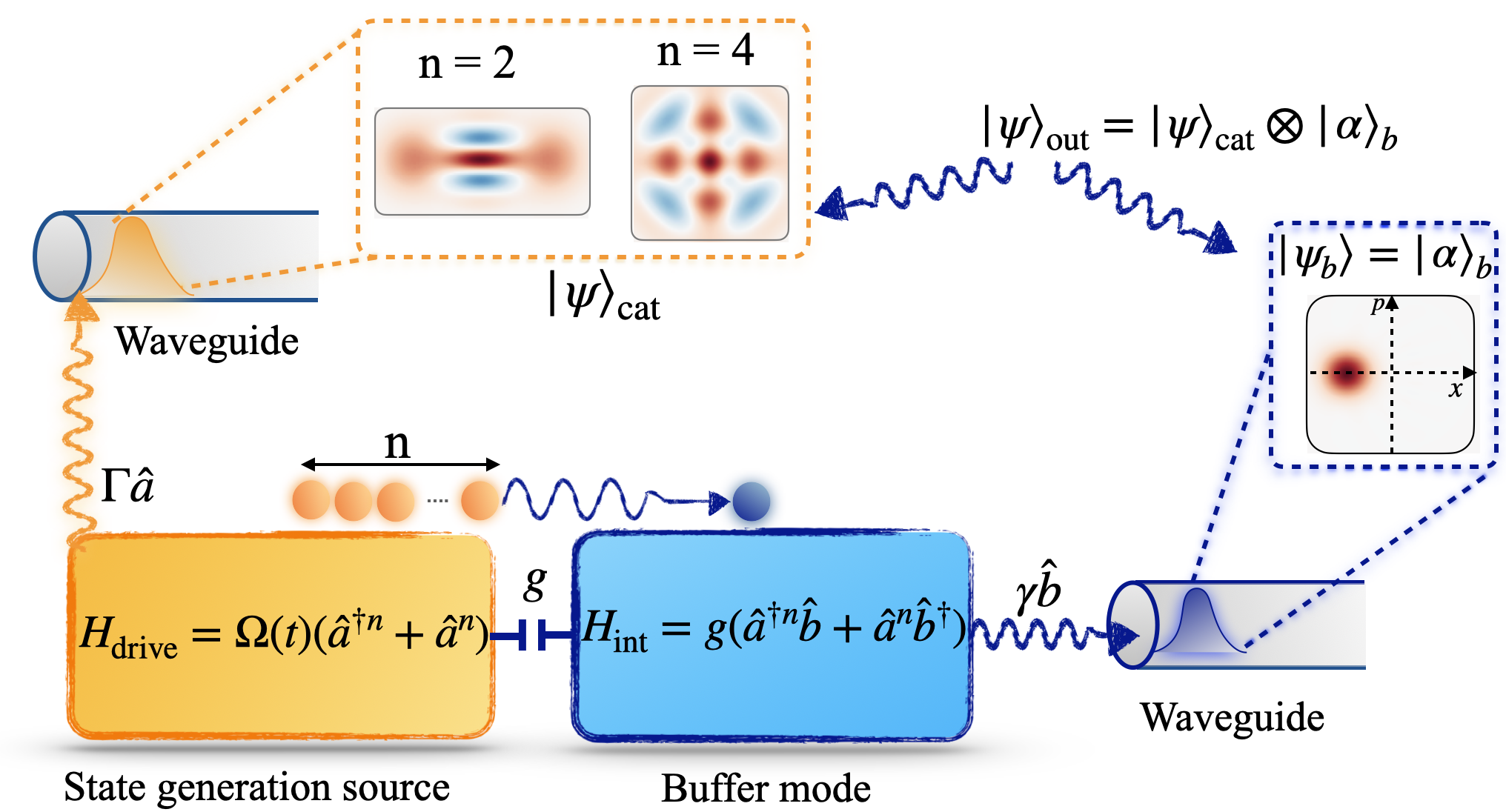}
\caption{Combination of engineered non-linear dissipation and linear dissipation for the preparation of traveling cat-state wave packets. The cat state is generated in the output field of the $a$-mode, state generation source (SGS) (orange box). The buffer $b$-mode (blue box) is coupled to the SGS through the \textit{$n$-to-$1$-photon} interaction $H_{\mathrm{int}}$. Both modes undergo strong linear transmission, $\Gamma$ and $\gamma$, into two separate waveguides. The total output field consists of a cat state $\ket{\psi}_{\mathrm{cat}}$ in the left waveguide and a coherent state $\ket{\alpha}_b$ in the right waveguide. The entire system is initialized in the vacuum state, and the time-dependent drive $H_{\mathrm{drive}}$ shapes the output fields.}
\label{Fig1}
\end{figure}
 
Unitary preparation of an $n$-legged cat state requires a nonlinear interaction proportional to $a^{\dagger n}a^n$ ($(2n)^{th}$ order of nonlinearity in field amplitude operators). A procedure based on such interactions has been analyzed in the Kerr-parametric oscillator \cite{nakamuracat}. As an alternative approach, one may obtain non-linear effects from an engineered dissipative coupling, $n$-photon decay, to the environment 
\cite{verstraete2009,PhysRevX.6.011022,kapit2017}. 
The use of dissipation to generate non-classical states and achieve steady-state entanglement in stationary modes has been proposed 
and demonstrated in various quantum systems \cite{PhysRevA.88.032317, PhysRevLettklaus}, and preparation of stationary bosonic states has been extensively studied both theoretically 
\cite{wolinsky1988, leghtas2015confining, minganti2016exact, PhysRevX.8.021005,  harrington2022engineered} 
and experimentally \cite{lescanne2020exponential,reglade2023quantum, marquet2024harnessing}. In this letter, we demonstrate that an engineered non-linear dissipation channel can affect a quantum bosonic system such that its linear emission into a waveguide forms high-fidelity propagating quantum states in single wave packet modes. Our theory thus marks a significant step toward the development of hardware-efficient quantum processors for long-distance communication.

\textit{Methods---}
We consider a single oscillator mode, described by the Lindblad master equation $\dot{\hat{\varrho}} =-i[\hat{H},\hat{\varrho}]+ \kappa\mathcal{D}(\hat{L})\hat{\varrho}$ where $\mathcal{D}(\hat{L})\hat{\varrho} = \hat{L}\hat{\varrho} \hat{L}^\dagger - \frac{1}{2}\{ \hat{L}^\dagger \hat{L},\hat{\varrho} \}$, $\kappa$ is a dissipation rate, and the Hamiltonian  $\hat{H}$ and Lindblad operator $L$ are expressed in terms of the ladder operators $(\hat{a},\hat{a}^\dagger)$. Considering the Hamiltonian $H = \Omega_d (\hat{L}+\hat{L}^\dagger)$, the master equation can be written in the compact form $ \dot{\hat{\varrho}} =  \kappa\mathcal{D}(\hat{L} - \lambda) \hat{\varrho}$ which has the stable steady state satisfying $\hat{L} \varrho_{\text{steady}} = \lambda \varrho_{\text{steady}}$ where $\lambda = e^{\frac{i3\pi}{2}}\frac{2\Omega_d}{\kappa}$ is the eigenvalue corresponding to the eigenstate of $\hat{L}$. A linearly driven and damped oscillator ($\hat{L}=\hat{a}$) has a coherent steady state, while assuming $\hat{L}=\hat{a}^2$ or $\hat{L}=\hat{a}^4$ leads to the steady-state generation of two- and four-component Schr\"odinger cat states, respectively \cite{Mirrahimi1,mirrahimi2016cat,mirrahimi2014dynamically}. This can be generalized to a two-mode Lindblad operator defined by $\{\hat{a}_1,\hat{a}_2\}$, with $\hat{L}=\hat{a}_1\hat{a}_2$, which stabilizes a pair-cat state \cite{agarwal1988nonclassical} and will be further explored at the end of the paper. 

In this Letter, we show that it is possible to employ engineered dissipation with a time-dependent drive $\Omega_d(t) =\Omega p(t)$, where $|p(t)|\leq1$ controls the shape and the parameter $\Omega$ specifies the strength of the drive, to prepare and release two- (four-) component cat states into traveling wave packets in one and the same process. To obtain the propagating wave packet quantum state, we permit a strong linear loss of the cavity field into a transmission waveguide, i.e., we supplement the master equation with an additional Lindblad damping term $\Gamma\mathcal{D}(\hat{a})\varrho $ with the corresponding constant loss rate $\Gamma$ of the same order as the driving amplitude $\Gamma \approx \Omega$.
As illustrated in Fig.\ref{Fig1}, an \textit{n-photon} loss $\hat{L}\propto \hat{a}^n$ is accomplished by engineering the interaction with a second quantum system, the so-called \textit{"buffer mode"} through the interaction Hamiltonian $\hat{H}_{\text{int}} =  g_{ab} (\hat{a}^n\hat{b}^\dagger+\hat{a}^{\dagger n}\hat{b})$; schematics shown in Fig. \ref{Fig1}. The buffer mode with field operators $\{\hat{b},\hat{b}^\dagger\}$ is strongly coupled to a waveguide through a linear loss Lindblad operator $\hat{L}_b = \gamma \hat{b},\,\gamma\gg g_{ab}$, leading to the Lindblad master equation of the combined system 
\begin{align} \label{method1}
   \dot{\varrho} = &-i\big[\Omega_d(t)(\hat{a}^n+\hat{a}^{\dagger n})+ g_{ab} (\hat{a}^n\hat{b}^\dagger+\hat{a}^{\dagger n}\hat{b}),\varrho\big] \nonumber\\&+\gamma\mathcal{D}(\hat{b})\varrho + \Gamma\mathcal{D}(\hat{a})\varrho.
\end{align}

We solve this master equation, and we show that for realistic parameters for superconducting circuit platforms, and an appropriately adjusted driving field $\Omega_d(t)$, the microwave field in the transmission waveguide coupled to mode $a$, indeed, populates a single wave packet mode cat state with high fidelity.
Henceforth, we refer to the $a$-mode system as the state generation source (SGS). It is important to mention that, since the photon decay rate is comparable to the drive amplitude, accelerating the generation of the cat state becomes important. This can be achieved by employing a fast and strong drive profile; however, such an approach may induce undesired transitions between different eigenstates of the superconducting circuit. To address this, incorporating a \textit{counter-adiabatic} drive term \cite{delcampo-shortcut}, slightly modifies the drive $\Omega(t)$ in Eq.(\ref{method1}) and can be effective in achieving higher-fidelity cat states; see the Supplemental Material (SM) for more details \cite{sup}.

\textit{Characteristics of the buffer mode---} To propagate a high-fidelity cat state, it is important to note that any possible entanglement of the cat state wave packet with the buffer mode output field decreases the fidelity; however, our simulation results show that the output field of the buffer mode is a pure coherent state, $\ket{\psi}_{\text{buff}}^{\text{out}} = \ket{\alpha}$, decoupled from the propagating cat state.

As the buffer mode has a high decay rate, $\gamma \gg g_{ab}$, it can be adiabatically eliminated:  $\dot{b} = -ig_{ab}a^n -\frac{\gamma}{2} b \approx 0 \Rightarrow b = \frac{-i2g_{ab}a^n}{\gamma}$. This results in the effective Lindblad master equation for the reduced system, $\dot{\varrho} =  \frac{4g_{ab}^2}{\gamma}\mathcal{D}(\hat{a}^n - \alpha^n(t)) \varrho + \Gamma\mathcal{D}(\hat{a})\varrho$, 
where $\alpha^n(t) =e^{\frac{i3\pi}{2}}\frac{\Omega_d(t)\gamma}{2g_{ab}^2}$. 
The engineered anti-commutator loss term in the master equation corresponds to the evolution by a non-Hermitian Hamiltonian $\hat{H} \propto i \hat{L}^\dagger \hat{L} \approx i \hat{a}^{\dagger n }\hat{a}^n$ and thus a $(2n)^{th}$-order nonlinearity term in field ampltitude operators, while we need only $(n+1)^{st}$-order interaction terms in the actual interaction between the SGS and the buffer mode. We note that, as the coherent state of the buffer mode is an eigenstate of the jump operator $\sqrt{\gamma}\hat{b}$, the effect of the buffer mode can be considered as equivalent to the separable no-jump evolution of the a-mode due to $\hat{H} \propto i \hat{L}^\dagger \hat{L}$. 

\textit{Extension to the full cat qubit basis ---} 
Starting from the vacuum state, the dissipative master equation \eqref{method1} with Lindblad operators $\hat{a}^2$ and $\hat{a}^4$ prepares 2-legged and 4-legged cat states with equal amplitudes on coherent states $\ket{C^{+2}_{\alpha}}=\mathcal{N}(\ket{\alpha} +\ket{-\alpha})$ and $\ket{C^{+4}_{\alpha}}=\mathcal{N}(\ket{\alpha}+\ket{-\alpha}+\ket{i\alpha}+\ket{-i\alpha})$, respectively. Defining these as logical qubit $|0\rangle$ states, we need a protocol to prepare also the respective logical qubit $|1\rangle$ states, $\ket{C^{-2}_{\alpha}}=\mathcal{N}(\ket{\alpha} -\ket{-\alpha})$ and $\ket{C^{-4}_{\alpha}}=\mathcal{N}(\ket{\alpha}+\ket{-\alpha}-(\ket{i\alpha}+\ket{-i\alpha}))$ and qubit superposition states. To this end, we employ a linear or parametric drive, 
$V = \epsilon\, p(t)\,(\hat a^n + \hat a^{\dagger n}), \, n=1(2)$, which executes a rotation in the 2(4)-legged cat qubit space because of the dissipative suppression of state components outside this space \cite{mirrahimi2014dynamically,Mirrahimi1}. The dissipative protection of the qubis space requires $|\epsilon| \ll |\alpha|^2 \kappa_n$, where $\kappa_n = 4g_{ab}^2/\gamma$; see \cite{mirrahimi2014dynamically,Mirrahimi1} for more details. In our numerical analysis, we assume the same drive profile as in Eq.~\eqref{method1}, with the additional requirement $\epsilon \ll \Omega$, and we find that the added drive does not alter the buffer mode output state, and we can produce high-fidelity cat qubit superposition states in outgoing single-mode wave packets. We show one realization of the cat qubit in the example section, and defer the possibility of further optimizing the drive profiles for future study.

\begin{figure}
\includegraphics[width=1.0\linewidth]{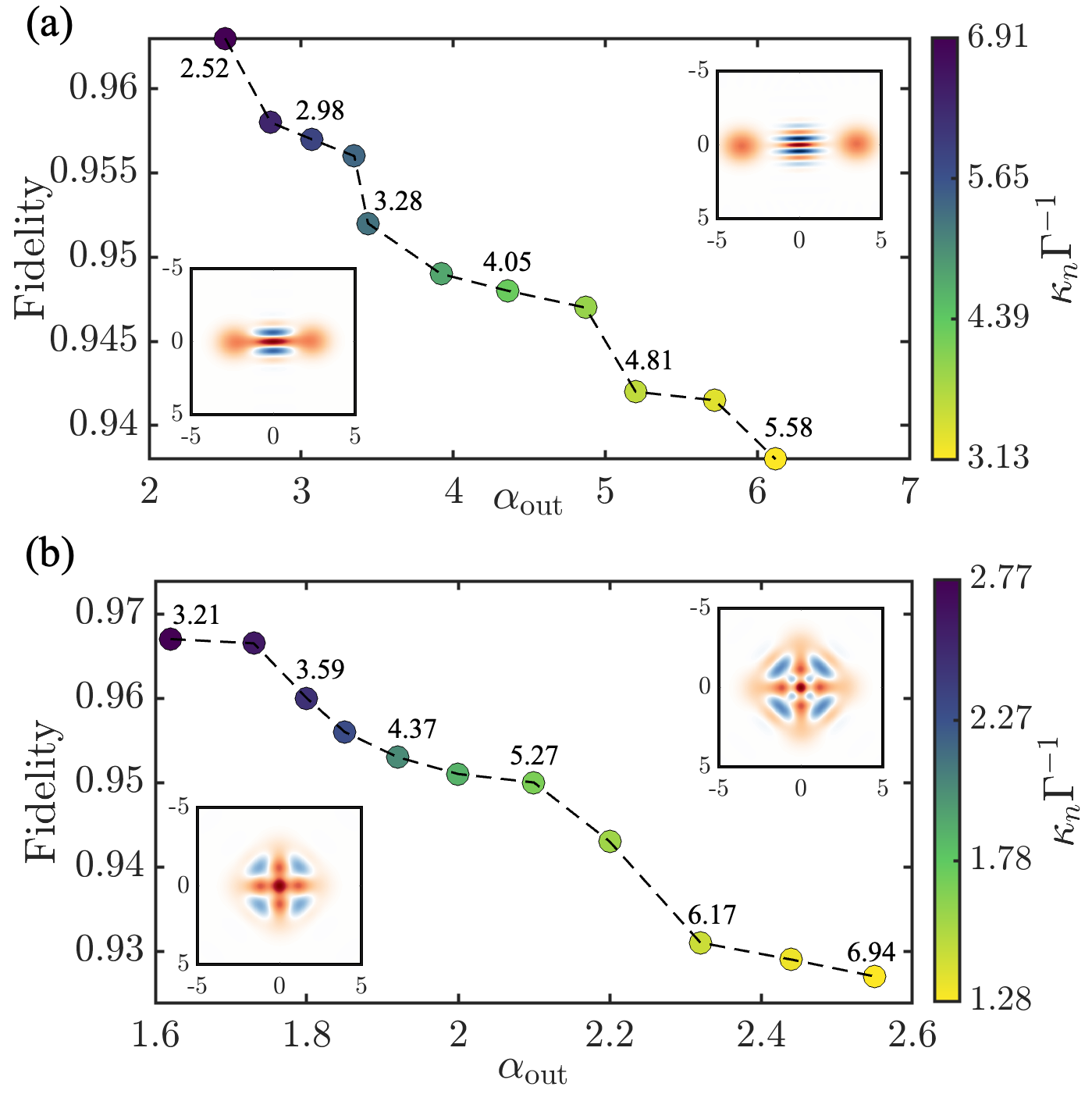}
\caption{The fidelity of the output field 2-legged and 4-legged cat states is shown together with the corresponding achievable amplitude parameters $\alpha_{\text{out}}$. The relative values of the coupling and dissipation rate, $\gamma/g_{ab}$,  are indicated by the number on several filled circles in each panel. The colorbar indicates the ratio between linear $\Gamma$ and nonlinear dissipation $\kappa_n$. The left and right insets show the 
Wigner function of the highest and lowest fidelity 2-legged and 4-legged cat states (here, and in all figures, the Wigner functions are rotated to align with the real and imaginary axes).
}
\label{fig3fidelity}
\end{figure}

\textit{Tradeoff between linear and nonlinear dissipation---} 
The SGS is subject to linear and non-linear dissipation channels with rates $\Gamma$ and $\kappa_n = 4g_{ab}^2/\gamma$, which must be properly balanced to produce the high-fidelity cat states. 
In Fig.~\ref{fig3fidelity}, we show the fidelity and size of cat states obtained for various combinations of decay rates and coupling strengths. In both panels, the color represents the ratio $\kappa_n/\Gamma$ {\it cf.} the right color bar, and the numbers indicated above the filled circles inside each panel box, correspond to $g_{ab}/\gamma$.  For each set of parameters, we identify the value of $\alpha =\alpha_{\text{out}}$ maximizing the state fidelity  $F=\bra{\psi}\varrho_{\text{out}}\ket{\psi}$ of the 2- and 4-legged cat state. In the resulting $(\alpha_{out},F)$ plot, we see that the fidelity is in the range of 95\% or more for cat sizes relevant for quantum communication and computing. Note also the insets in the figure showing the Wigner functions of the smallest and largest cat states.   

\textit{Experimental proposal---} Controlling the interaction between microwave photons in superconducting circuits relies on Josephson junction (JJ) elements to implement non-linear dynamics through the potential $U(\hat{\varphi}) \propto E_J (1- \cos(\hat{\varphi}))$ where $\hat{\varphi}$ is the phase across the junction with the junction energy $E_J$ \cite{devoret2005implementing}. The JJ acts as a nonlinear inductance, and its response can be tuned \textit{in situ} by applying a phase bias, which is achieved by threading a magnetic flux through a superconducting loop including the JJs. In general, the potential of a loop of multiple JJs can be written as $U(\hat{\varphi},\Phi) = \sum\limits_{m=2}^\infty \frac{C_m(\Phi)}{m!} (\hat{\varphi} - \varphi_0)^m$, where the coefficients $C_m(\Phi) = \partial^m U/\partial \varphi^m |_{\varphi=\varphi_0}$ yield the Taylor expansion around $\varphi_0$ minimizing the potential and depends on the magnetic flux $\Phi$ threaded through the loop. The flux drive $\Phi$ can be tuned to implement a particular combination of linear and nonlinear interactions and to avoid escape to unconfined states \cite{miano2022frequency,PhysRevApplied.11.014030}. The asymmetrically
threaded SQUID (ATS) \cite{lescanne2020exponential} consists of a SQUID (superconducting quantum interference device), including two JJs with junction energies $E_1, E_2$ in parallel, shunted in the center by a large inductance $L_J$. This device comprises two loops with the corresponding flux drives $\varphi_1,\varphi_2$, respectively; See Sec. A in \cite{sup}. 

Without loss of generality, we consider a symmetric SQUID, i.e. $E_1=E_2 = E_J$, with the potential of the ATS obtained as $-\hat{U}(\hat{\varphi}) =  -\frac{\hat{\varphi^2}}{2L_J} + 2E_{J}\cos(\varphi_\Sigma) \cos(\hat{\varphi}+\varphi_\Delta)$ where $2\varphi_\Sigma = \varphi_1+\varphi_2,\, 2\varphi_\Delta=\varphi_1-\varphi_2 $.
To suppress the dominant self-Kerr and cross Kerr resonant interactions, we assume the difference between the two dc bias drives is $\varphi_{\Delta} = \pi/2$. Then, the Hamiltonian of the ATS depends only on the nonlinear odd terms of the flux operator $ H \propto \sin(\hat{\varphi}) \propto \sum_k\hat{\varphi}^{2k+1} $. Finally, the flux drive ${\varphi}_{\sum} $ is adjusted to implement the interaction needed.

The full circuit follows the ATS design for both SGS and buffer mode, which are capacitively coupled with the interaction coupling $g_{ab}$. 
Introducing the dressed mode operators $(\hat{\bm{a}},\hat{\bm{b}})$ with frequencies \((\bm{\omega}_a,\bm{\omega}_b)\) corresponding to the SGS and buffer modes, respectively, the effective Hamiltonian is obtained as $\hat{\bm{H}} = \bm{\omega}_a \bm{\hat{a}}^\dagger\bm{\hat{a}} + \bm{\omega}_b \bm{\hat{b}}^\dagger\bm{\hat{b}} + \sum\limits_{k=0}\mathcal{C}_k^a\big[\varphi_a \bm{\hat{a}}+\varphi_b \bm{\hat{b}}+h.c.\big]^{2k+1}+ \sum\limits_{{k=0}}\mathcal{C}_k^b\big[\varphi'_a \bm{\hat{a}}+\varphi'_b \bm{\hat{b}}+h.c.\big]^{2k+1}$ where the coefficients $(\varphi_{a,b},\varphi'_{a,b},\mathcal{C}_k^{a,b})$ are evaluated in \cite{sup}. By adjusting a flux drive on the buffer mode with the frequency $\propto n\bm{\omega}_a-\bm{\omega}_b$ and introducing an $n$-photon drive on the SGS through the charge line $\propto [ \bm{\hat{a}}^\dagger e^{in\bm{\omega}_at}+h.c]$, in the rotating frame of $\{\bm{\omega}_{a},\bm{\omega}_b\}$, the effective Hamiltonian is evaluated as
$\hat{H}_{\mathrm{eff}}  = \Omega_{n}(t)(\bm{\hat{a}}^n +\bm{\hat{a}}^{\dagger n})+g_n(\bm{\hat{a}}^n\bm{\hat{b}}^\dagger + \bm{\hat{a}}^{\dagger n}\bm{\hat{b}}), 
$ where the coefficients $\Omega_n(t),g_n$ depend on the circuit parameters \cite{sup}. By coupling the SGS and buffer mode to the waveguides, the dynamics of the quantum circuit can be described by the evolution in Eq. \eqref{method1}.
It is worth noting that, compared to recent papers on stabilizing the two-legged cat state \cite{mirrahimi2016cat,leghtas2015confining,lescanne2020exponential,reglade2023quantum}, they consider a linear resonator as a storage cavity and apply both drives on the buffer mode. However, in our proposal, we use a separate ATS for the SGS and apply only the n-photon decay to the buffer mode for two important reasons. First, this prevents the transition of photons from the b-mode to the SGS and then into the waveguide, ensuring that the output field only populates a single-mode wavepacket. Second, since we generate and release the state in the same process,  we do not rely on long storage and coherence times of the resonator. 

\begin{figure}[t]
\includegraphics[width=.49\textwidth]{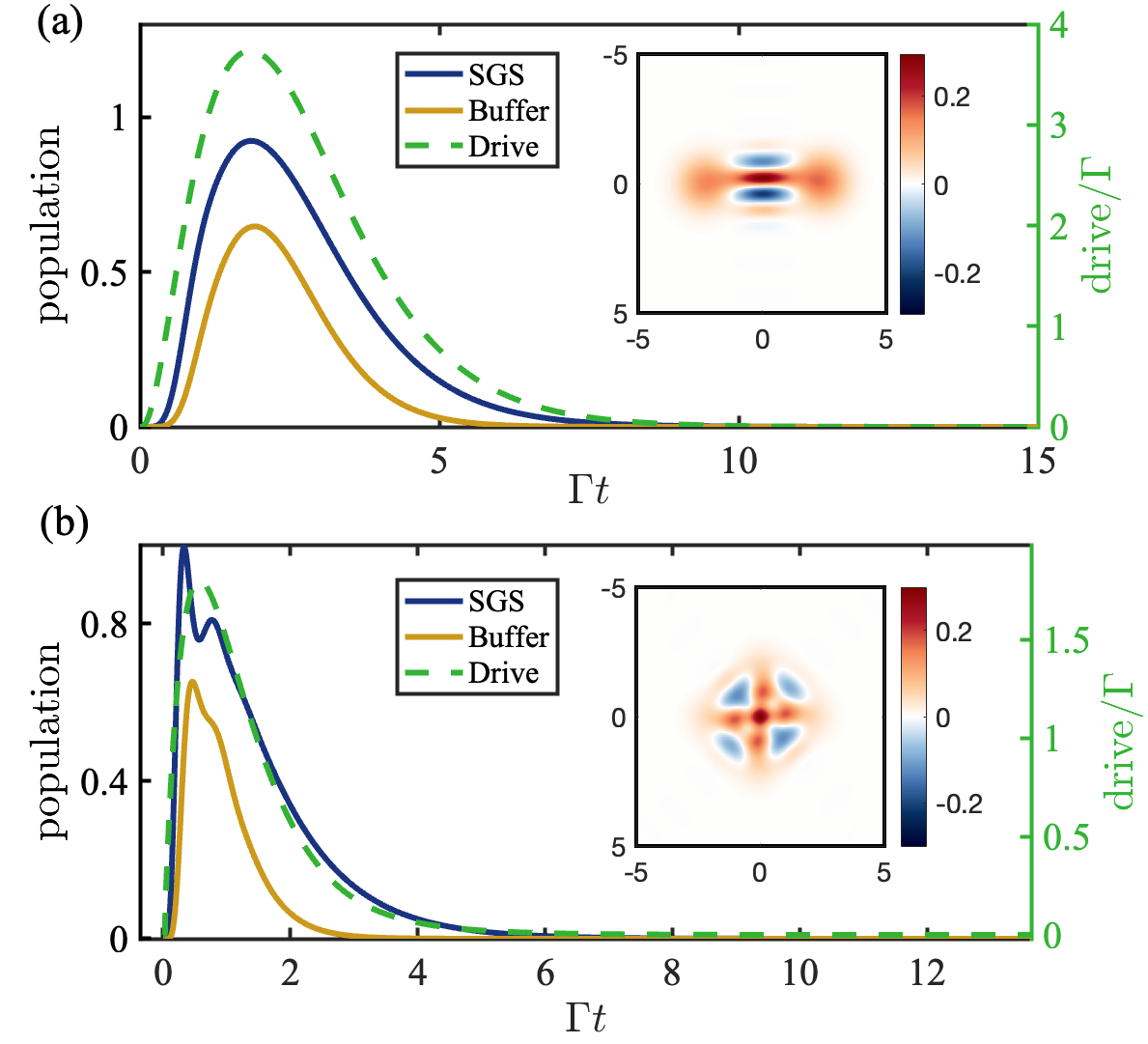}
\caption{ Panels (a) and (b) show the population of the SGS and the buffer mode (solid curves) on the left y-axis, and the total drive amplitude  $(\Omega_d+\Omega_{ca})/\Gamma$ (dashed curve) on the right y-axis, corresponding to 2-cat and 4-cat generation, respectively. The inset plots in (a) and (b) illustrate the Wigner functions corresponding to propagating 2-cat and 4-cat qubit states obtained by the simulation parameters $\kappa_n/\Gamma = 6.9,2.0$ and rotating drive strength $\epsilon/\Omega= 5\%,8\%$, respectively; see the text for more details.}
\label{Fig3}
\end{figure}
\textit{Example---} To obtain the quantum state of the propagating mode, we utilize the master equation \eqref{method1} and the quantum regression theorem to evaluate the two-time correlation function of the emitted field and its mode decomposition $\mathcal{G}^{(1)}(t_1,t_2)=\kappa\langle \hat{a}^\dagger(t_1)\hat{a}(t_2) \rangle = \sum\limits_i n_i v_i^*(t_1)v_i(t_2)$\cite{PhysRevLett.123.123604}, where $\{v_i(t)\}$ are orthonormal temporal modes of the output field with the corresponding mean photon number $n_i$; see the End Matter for more details. The aim is for the output field to populate only one mode, $v_1$, with a mean photon number close to the total output, $n_1 \approx n_{\text{out}}$. 

Fig. \ref{Fig3} illustrates one realization of the SGS and buffer evolution. Panels (a) and (b) show the SGS and buffer population during the state generation process and the drive profile corresponding to the 2-legged and 4-legged cat states, respectively. The emission rate effectively suppresses the instantaneous population of higher Fock states inside the SGS, while over time, several photons are emitted and populate the traveling wave packet. 
The inset panels in Fig.~\ref{Fig3}(a,b) show the Wigner functions of the most populated mode, revealing the expected two-legged and four-legged cat qubit states, respectively. Considering the  the logical basis $\{\ket{C^{+n}_{\alpha}},\ \ket{C^{-n}_{\alpha}}\}$ and probabilities $\{p_+, p_-\}$, a logical qubit superpostion state can be expressed as $\sqrt{p_+}\ket{C^{+n}_{\alpha}} + e^{i\phi}\sqrt{p_-}\ket{C^{-n}_{\alpha}}$, for both 2-cat and 4-cat qubits. In Fig.~\ref{Fig3}(a), we find a fidelity of $95.3\%$ with the cat-state qubit for parameters $p_+ = 0.75$, $p_- = 0.25$, $\phi = -0.8$, and $|\alpha|^2 = 2.52$, with $96\%$ of the population in a single mode. In Fig.~\ref{Fig3}(b), the corresponding parameters are $p_+ = 0.9$, $p_- = 0.1$, $\phi = -0.4$, and $|\alpha|^2 = 1.85$, yielding a fidelity of $94.5\%$ and a mode population ratio of $95\%$ in mode $v_1$. It is important to note that the two-legged cat state can be generated faster than the four-legged cat state, as it only requires third-order nonlinearity. Considering values of $ 1/\Gamma = 0.22,\, 1.32~\mu\text{s} $, the total generation times obtained from Fig.~\ref{Fig3} are approximately $T \approx 2~\mu\text{s} $ and $ 9.3~\mu\text{s} $ for the two-cat and four-cat states, respectively; see SM \cite{sup} for more details.

In addition, we analyze the evolution of the buffer mode over the entire duration of the state generation processes and find that the buffer mode output has fidelity of $99.0\%$ and $98.5\%$, with single mode coherent quantum states $ \ket{\psi}_{\text{buffer}} = \ket{\alpha = -3.85 }$ and $ \ket{\psi}_{\text{buffer}} = \ket{\alpha = -2.39} $, corresponding to the 2-legged and 4-legged cat state generated by the SGS. As mentioned earlier, the coherent state of the buffer-mode output field leads to a high-fidelity propagating cat state, as shown in Fig. \ref{Fig3}. 

\textit{Summary---}
We have demonstrated a deterministic and hardware-efficient method for generating propagating Schrödinger cat states using superconducting circuits. By engineering two- and four-photon loss, we have shown that it is possible to produce high-fidelity two- and four-component cat states in single traveling wave packet modes. The driving pulse, together with the nonlinear coupling to the lossy buffer mode, controls the formation of the quantum state as it is gradually released to the waveguide. We have shown that by applying a proper linear or quadratic drive with a time-dependent profile, simultaneously with the $n$-photon drive, we can also produce high-fidelity cat qubit superposition states, which are essential for applications in quantum communication and computing. 

Our calculations show that the emission from the buffer mode is well approximated by a single wave packet mode coherent state. This implies that the system is always in an eigenstate of the buffer mode jump operator $\sqrt\gamma \hat{b}$, and hence the state evolution is governed exclusively by the no-jump dynamics, cf. the non-Hermitian Hamiltonian. This is why we benefit directly from the $(2n)^{th}$ order non-Hermitian term $\propto i\hat{a}^{\dagger n}\hat{a}^n$, arising from the lower order interaction $\propto (\hat{a}^n \hat{b}^\dagger + \hat{a}^{\dagger n} \hat{b})$ to the buffer mode.         
As an important application of our proposal, we studied the generation of propagating grid states from propagating 4-cat states, marking a significant step toward large-scale fault-tolerant quantum communication and computing. As detailed in the Supplemental Material, our theory extends to the generation of entangled propagating states, such as pair-cat states. We further note that propagating non-Gaussian quantum states may also serve as sensitive probes for metrological purposes \cite{PhysRevA.95.012305,maryam1,PRXQuantum.4.020337,GKP}.
The state generation source and the buffer mode may both be subjected to additional
noises, e.g., dephasing and other dissipation channels; however, our approach
inherently mitigates such noises due to the simultaneous generation
and release of the quantum states, which provides a significant advantage over the stationary state
preparation methods.

Looking ahead, utilizing optimal control may enable faster preparation and higher-fidelity cat qubit states.  Addressing and controlling higher-order terms in the expansion of the Josephson nonlinearities may improve the cat- and grid-state generation beyond our analysis, relying on low-order approximations. 


\textit{Authors' contributions---} M.K. developed the original idea and the experimental proposal and carried out the theoretical simulations. Both authors contributed to the analyses of the results and to the writing of the article.

\textit{Acknowledgment---} M.K. acknowledges fruitful discussions with G. Johansson and useful comments from S. Girvin and T. Hillmann. 
This work was supported by the Knut and Alice Wallenberg Foundation through the Wallenberg Center for Quantum Technology (WACQT) and the Danish National Research Foundation Center for Quantum Hybrid Networks (DNRF 139). The numerical simulations were performed using resources provided by the Chalmers Center for Computational Science and Engineering (C3SE) under project 2025/1-6.

\bibliographystyle{apsrev4-2}
%

\newpage
\section{End Matter}
\subsection{1- Catching SGS output field}
To investigate the output field from the SGS, we simulate Eq. \eqref{method1} and use the quantum regression theorem to evaluate the two-time correlation function $\mathcal{G}^{(1)}(t_1,t_2)=\kappa\langle a^\dagger(t_1)a(t_2) \rangle$. We next identify the mode decomposition $\mathcal{G}^{(1)}(t_1,t_2)  = \sum\limits_i n_i v_i^*(t_1)v_i(t_2)$, where $\{v_i(t)\}$ are orthonormal temporal modes of the output field with corresponding mean photon number $n_i$. The aim is that the output field of the SGS populates only one mode $v_1$, with a mean photon number close to the total number of photons in the output field, $n_1 \approx n_{\text{out}}$. To obtain the quantum state of the output field, we employ \cite{PhysRevLett.123.123604} and simulate the state of the field building up in an artificial downstream cavity with ladder operators $\{\hat{d},\hat{d}^\dagger\}$. Such a cavity will fully capture the quantum state contents of the mode $v_1$, if the coupling to the wave guide is, $g_{v_1}(t) = -{v_1}^*(t)/\sqrt{\int_0^t |v_1(t)|^2}$, and we hence recover the state of the wave packet mode by solving the cascaded systems master equation including the time-dependent coupling interaction $H_{da}(t) = i\frac{\sqrt{\Gamma}}{2}( g_{v_1}^*(t)\hat{a}^\dagger \hat{d}-g_{v_1}(t) \hat{d}^\dagger \hat{a}) $ and the Lindblad operator $\hat{L}_{sv} =\sqrt{\Gamma} \hat{a}+ g_{v_1}^*(t) \hat{d}$ \cite{PhysRevLett.123.123604,combes2017slh}.

\subsection{2- Application: Propagating Grid States}  
\begin{figure}[ht!]
\includegraphics[width=1.\linewidth]{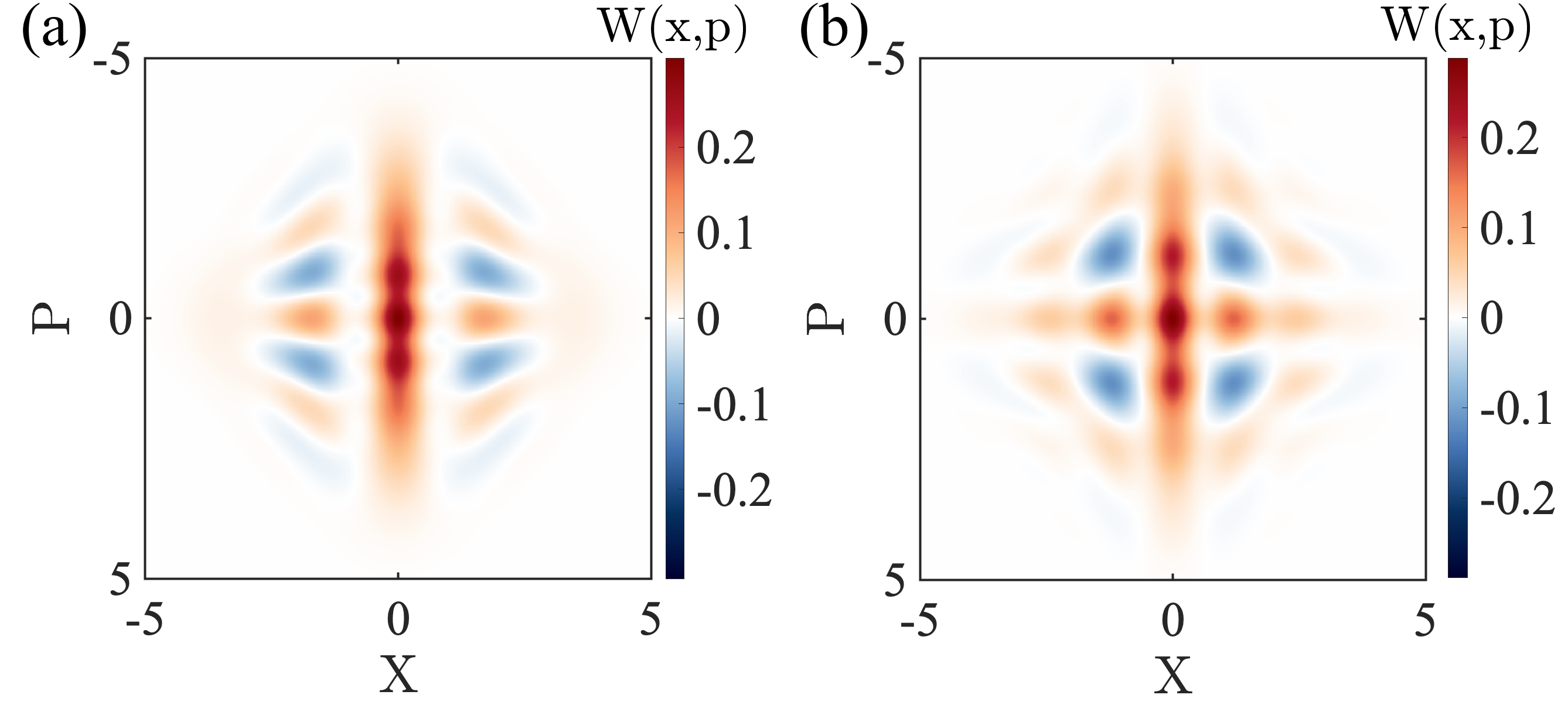}
\caption{Panel (a) and (b)  show the Wigner distribution of the propagating grid state obtained by the first and second iteration of the breeding protocol on the generated 4-cat state with fidelity 95\% with $|\alpha|^2 \approx$2.1, respectively.  The effective squeezing parameters \(\Delta_x = 3.6\, \mathrm{dB}\) and \(\Delta_p = 1.1\, \mathrm{dB}\) along the \(x\)- and \(p\)-axes, respectively, are evaluated for the panel (b). }
\label{figwigner}
\end{figure}
A significant example of propagating non-Gaussian quantum states is the traveling grid state, which serves as a sensitive sensor for sensing purposes and also as an important resource for obtaining fault-tolerant quantum computing and error correction 
\cite{GKP,terhalGKP,rafaelgkp,vuillot2019quantum,campagne2020quantum,terhal2020towards,PRXQuantum.2.030325,PRXQuantum.2.020101,noh2022low}.  
Our theory enables the breeding of wave packet quantum states, which can be combined on beam splitters and subjected to measurements of $x$ or $p$ quadratures that herald the presence of grid states in definite temporal modes in the unmeasured output port. While \cite{albert2018performance,PhysRevA.108.012603} suggest preparing grid state by breeding from binomial states $\propto\ket{0}+\ket{4}$, we breed from 4-legged cat state with dominant $\ket{0}$ and $\ket{4}$ components when $|\alpha|^2\approx 2.1$. The Wigner function after one and two iterations of the beam splitter and quadrature measurement on one output port are shown in Fig. \ref{figwigner} (a) and (b), see \cite{sup} for further details.

\subsection{3 - Extension to entangled-state wavepackets: generation of pair-cat states}  
Our theoretical approach can be generalized for the preparation of multi-mode entangled states, particularly pair cat states \cite{agarwal1988nonclassical,gou1996vibrational,albert2019pair,paircat2,gertler2023experimental}, which occupy wave packets traveling in different waveguides or different frequencies. For this to work, the wave packets are released from two distinct resonator modes with field operators \(\{\hat{a}_1, \hat{a}_2\}\) and resonance frequencies \(\{\bm{\omega_{a_1}}, \bm{\omega_{a_2}}\}\), respectively; See Fig. 3 in \cite{sup}. These modes are both coupled to the SGS and buffer mode of Fig. \ref{Fig1}, causing correlated losses 
described by the  master equation 
$ \dot{\hat{\varrho}} =  \kappa\mathcal{D}(\hat{L} - \lambda) \hat{\varrho}$, with $\hat{L}\propto \hat{a}_1^2 \hat{a}_2^2 $. The steady-state solutions of this equation are the so-called pair-coherent states \( \ket{\alpha,\alpha}_{a_1,a_2}\propto\sum_n \frac{\alpha^{2n}}{n!}\ket{n, n}
\), and superposition states such as the  \textit{pair-cat state} with only even numbered Fock state components,
\(
\ket{\text{pair-cat}} \propto \ket{\alpha,\alpha}_{a_1,a_2} + \ket{i\alpha,i\alpha}_{a_1,a_2}\) \cite{sup}. These states can protect effective two-qubit entanglement and exponentially suppress dephasing errors and arbitrary photon loss in both modes \cite{albert2019pair}.

Implementation of the joint Lindblad operator \(\hat{L}_{12} \propto \hat{a}_1^2 \hat{a}_2^2\) can be achieved by a flux drive on 
the buffer mode with a frequency \(\omega_d=\bm{\omega_b} - 2(\bm{\omega_{a_1}} + \bm{\omega_{a_2}})\) and on the SGS with the frequency $\propto 2(\bm{\omega_{a_1}} + \bm{\omega_{a_2}})$ which lead to the effective Hamiltonian 
\(H_{\text{eff}}^{\text{pair}} = \Omega_d(t)(\hat{a}_1^2\hat{a}_2^2 + h.c.) + g(\hat{a}_1^2\hat{a}_2^2 b^\dagger + h.c.)\); see the details in \cite{sup} Sec. D.
To simultaneously release the entangled state into wave packet modes in two wave guides, we consider the same constant decay rate for both modes, 
\(L_{a_1} = \sqrt{\Gamma}\hat{a}_1, L_{a_2} = \sqrt{\Gamma}\hat{a}_2\), \textit{i.e.} in Eq.\eqref{method1} the SGS dissipation changes as $\Gamma\mathcal{D}(\hat{a})\varrho\rightarrow \Gamma\mathcal{D}(\hat{a}_2)\varrho+\Gamma\mathcal{D}(\hat{a}_1)\varrho$. 
We assume the same drive profile as in Fig.\ref{Fig2-6panels}(b), with the parameter 
ratio \(\gamma/g_{ab} =7.2,\, 4g_{ab}^2/\Gamma \gamma =1.2\), and the total duration \(T = 6/\Gamma\). \(96\%\) of the output fields populate single modes, having the fidelity  \(F = 95\%\) with the pair-cat state with \(|\alpha|^2 = 0.94\). This high fidelity is promising for the sharing of entanglement and information between distant quantum processors.  

\end{document}


\title{Supplemental Material for "Environment-Assisted Generation of Non-Gaussian Wavepacket Quantum States"}
\author{Maryam Khanahmadi}
\email{m.khanahmadi@chalmers.se}
\affiliation{Department of Microtechnology and Nanoscience, Chalmers University of Technology, 412 96 Gothenburg, Sweden}
\author{Klaus M{\o}lmer}
\affiliation{Niels Bohr Institute, University of Copenhagen, Jagtvej 155A, DK-2200 Copenhagen,Denmark} 
\maketitle
\appendix

 \section{Details of the state generation source}\label{sec:ATS}
\begin{figure*}[ht!]
\includegraphics[width=.4\textwidth]{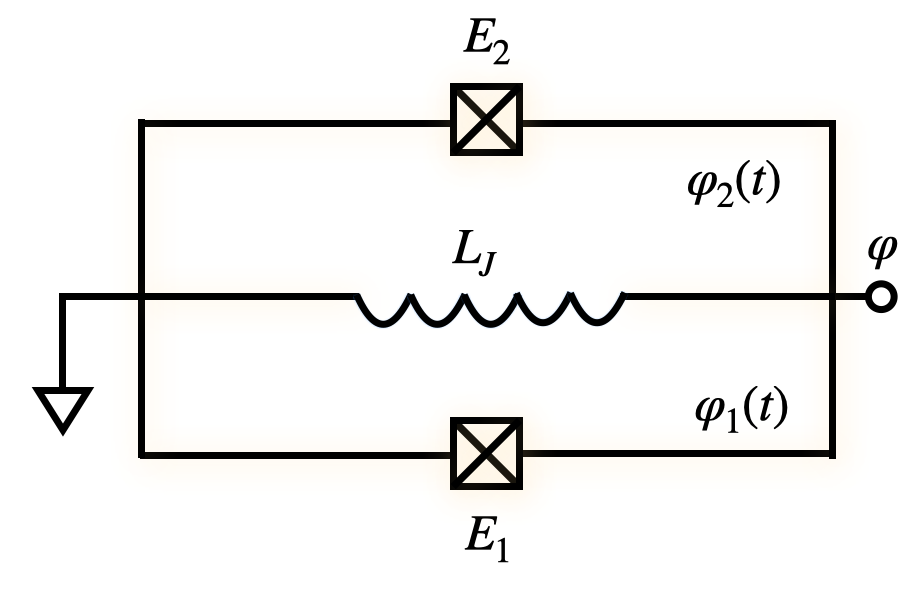}
\caption{Schematic of an asymmetry-threaded SQUID (ATS) \cite{SMlescanne2020exponential}. 
The ATS consists of two loops with external flux drives, $\varphi_1$ and $\varphi_2$, enabling control of the Kerr effect and higher-order nonlinear interactions. 
See Sec.~\ref{sec:ATS} for more details.
}
\label{ATS}
\end{figure*}

For the source of n-component cat state generation, we consider a circuit design with tunability of the nonlinear terms to achieve dominant $3^{rd}$ and $5^{th}$ order interactions in bosonic field amplitudes. Our main circuit component is an asymmetric-threaded SQUID (ATS) as shown in Fig. \ref{ATS}. The ATS includes two loops with junction energies $E_1, E_2$, in parallel to an inductance $L_J$, where each loop is affected by an external flux drive $\{\varphi_1,\varphi_2\} $\cite{SMlescanne2020exponential}. The dynamics of the system is governed by the Hamiltonian,
\begin{align}
    H_{\mathrm{ATS}} = 4E_c\hat{n}^2 + \hat{U}(\hat{\varphi}),
    \end{align}
where $E_c$ is the energy of the shunted capacitance, not shown in the figure \ref{ATS}, and
   \begin{align}\label{E1} 
    -\hat{U}(\hat{\varphi}) = E_1 \cos(\hat{\varphi} + \varphi_1) +E_2 \cos(\hat{\varphi} - \varphi_2)- \frac{\hat{\varphi}^2}{2L_J},
\end{align}
where $\hat{n},\hat{\varphi}$ correspond to the charge and flux operators, respectively. We consider $E_1 = E_2 =E_J$ and the first two terms of the potential \eqref{E1} can then be written as
\begin{align}
E_J [\cos(\hat{\varphi }+ \varphi_1)+ \cos(\hat{\varphi} - \varphi_2)] = 2 E_J \Bigg[\cos(\frac{2\hat{\varphi} + (\varphi_1-\varphi_2)}{2})\cos(\frac{\varphi_1+\varphi_2}{2})\Bigg],
\end{align}
where by introducing the new variables
\begin{align}
     \varphi_\Sigma = \frac{\varphi_1+\varphi_2}{2},
    \varphi_\Delta = \frac{\varphi_1-\varphi_2}{2},
\end{align}
the potential  can be written, 
\begin{align}\label{E2}
     -U(\hat{\varphi}) = - \frac{\hat{\varphi}^2}{2L_J} + 2E_{J} \cos(\varphi_\Sigma) \cos(\hat{\varphi}+\varphi_\Delta). 
 \end{align}
Defining ladder operators $\{\hat{a},\hat{a}^\dagger\}$ through the relations $\hat{\varphi}\propto(\hat{a}^\dagger+\hat{a}), \hat{n}\propto i(\hat{a}^\dagger - \hat{a})$, the linear part of the circuit specifies the oscillator form  $4E_c\hat{n}^2 + \hat{\varphi}^2/2L_J \equiv \omega_a \hat{a}^\dagger \hat{a}$, where $\omega_a = \sqrt{\frac{8E_c}{L_J}}$ (henceforth we set $\hbar =1$). To obtain leading $3^{rd}$ and $5^{th}$ order nonlinearities we suppress the even order Kerr effect by applying a DC bias $\varphi_\Delta = \pi/2$ and an RF drive $\varphi_\Sigma = \varphi_\Sigma^{dc} + \eta \cos(\omega_d t)$ leading to the potential
\begin{align}\label{ats}
    -\frac{U(\hat{\varphi})}{E_J} = 2 \cos(\varphi_\Sigma^{dc} + \eta \cos(\omega_d t)) \sin(\hat{\varphi}) =  2 \cos(\varphi_\Sigma^{dc} + \eta \cos(\omega_d t))\sum_{k}^\infty \frac{\hat{\varphi}^{2k+1}}{(2k+1)!},
\end{align}
where $\eta, \omega_d$ correspond to the flux drive's amplitude and frequency, respectively. 

Our design is composed of a separate state generating system (SGS) and engineered environment (EE). These are both ATS systems as sketched above, capacitively coupled to each other and    individually coupled to separate waveguides. The ATS design suppresses the resonant self Kerr and cross Kerr terms $\propto a^{\dagger n}a^n b^{\dagger m}b^m,n,m\ge1$ between the SGS and EE while permitting non-linear classical driving of the SGS and a multi-photon loss process through their $\hat{a}^n \hat{b}^\dagger$ interactions. To apply the n-photon drive on the SGS, utilizing Eq. \eqref{ats}, we consider $\varphi_\Sigma=\varphi_\Sigma^{dc}$ and a coherent drive through the charge line  $\Omega_d = e^{in\omega_a}a^\dagger + h.c.; n=2,4$. We also apply a flux drive $ \varphi_\Sigma^{b}=\pi/2 + \eta_b \cos(\omega_d^b t)$ on the "buffer mode" ATS circuit described by operators $\{b^\dagger,b\}$ with the corresponding frequency $\omega_b$. Assuming a weak flux drive amplitude $\eta_b\ll1$, the potential of the buffer mode is
\begin{align}
    -\frac{U(\varphi_b)}{E_J} = 2 \sin( \eta_b \cos(\omega_d^b t)) \sin(\hat{\varphi}_b) =  2 \eta_b \cos(\omega_d^b t)\sum_{k}^\infty \frac{\hat{\varphi}_b^{2k+1}}{(2k+1)!}.
\end{align}
By applying the buffer mode drive frequency $\omega_d^b = n\omega_a-\omega_b, n=2,4$, we obtain the interaction Hamiltonian $H_{\text{int}} \propto a^nb^\dagger + h.c.$, to be elaborated in more detail in the next section.

\section{Engineered-environment with a buffer mode}\label{sec:buffer}
\begin{figure*}[ht!]
\textbf{(a)\quad\quad\quad\quad\quad\quad\quad\quad\quad\quad\quad\quad\quad\quad\quad\quad\quad\quad\quad\quad\quad\quad\quad\quad\quad\quad\quad\quad\quad(b)}\par\medskip
\includegraphics[width=.5\textwidth]{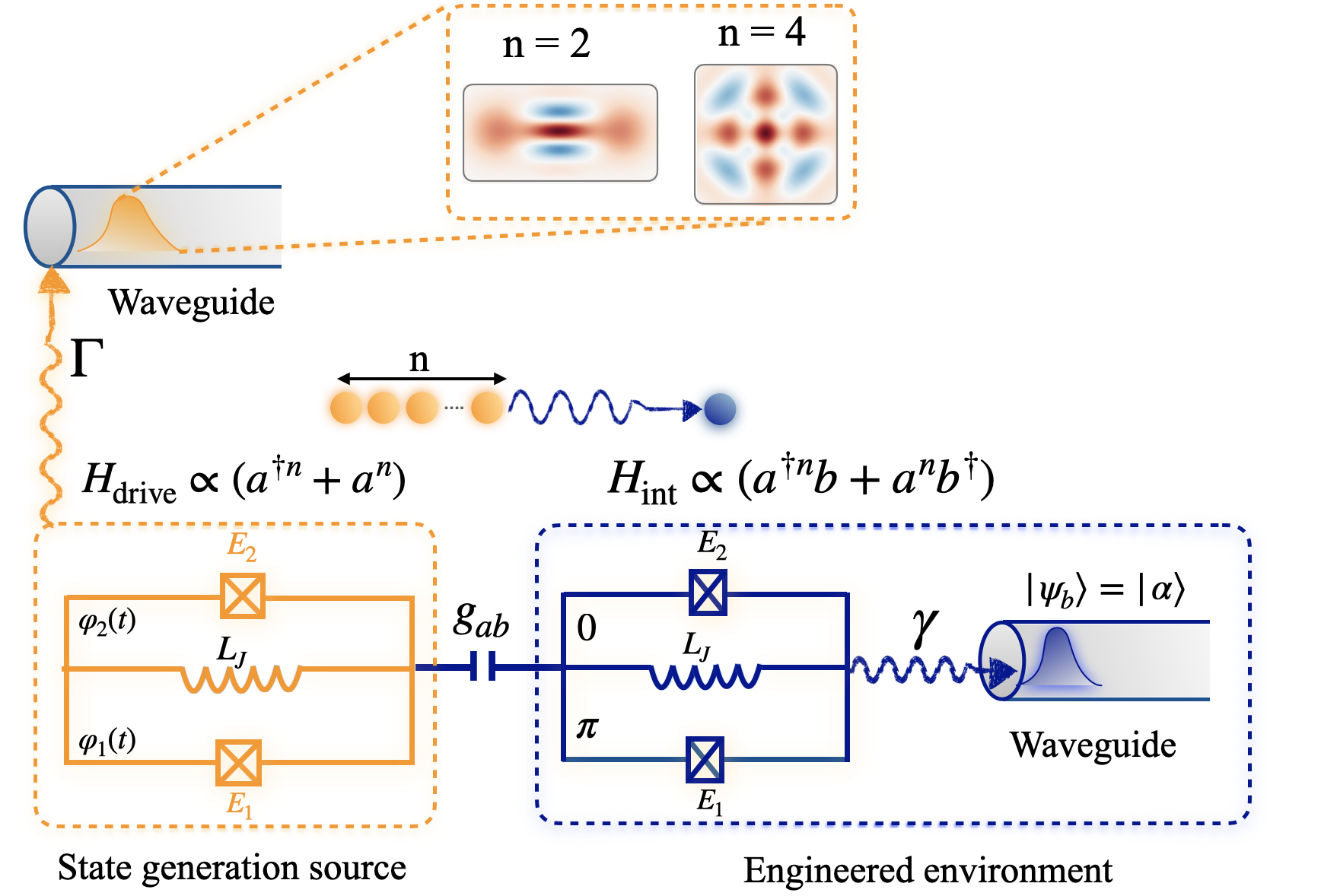}
\includegraphics[width=.43\textwidth]{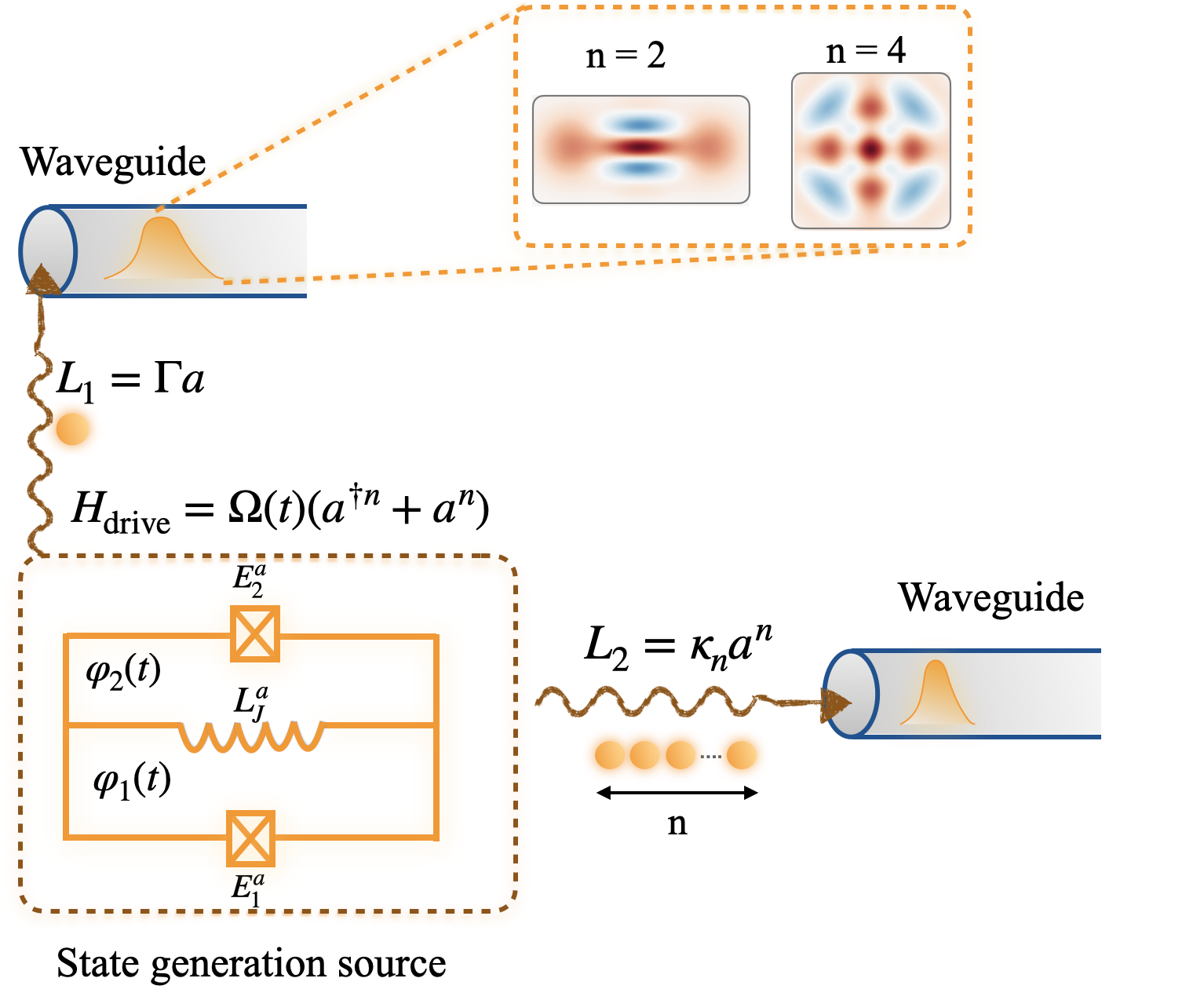}
\caption{ Schematic of the engineered-environment scenario. 
(a) The leaky buffer mode, characterized by the operators $\{\hat{b}, \hat{b}^\dagger\}$, interacts with the quantum state generation source (SGS) via the interaction Hamiltonian $H_{\mathrm{int}} \propto (\hat{a}^{\dagger n}\hat{b} + \hat{a}^n \hat{b}^\dagger)$. Both SGS and buffer mode are considered as the ATS design introduced in Sec. \ref{sec:ATS}. .
(b) Adiabatic elimination of the buffer mode induces an effective $n$-photon decay process on the SGS. Applying an $n$-photon drive on the SGS, $H_{\mathrm{drive}} = \Omega(t) (\hat{a}^n + \hat{a}^{\dagger n})$, leads to the generation of the desired 2-cat and 4-cat states through the single-photon decay channel.
}
\label{circuitshape}
\end{figure*}

According to Fig. \ref{circuitshape} (a), we consider the frequency $\{\omega_a,\omega_b\}$ for SGS and the buffer mode, respectively. Assuming these two modes are capacitively connected together $g(\hat{a}^\dagger \hat{b}+\hat{b}^\dagger \hat{a})$, the total Hamiltonian of the circuit is written as a combination of a linear and non-linear terms as follows
\begin{align}
    H_{\mathrm{circuit}} = H_{L} + H_{nL} = &\omega_a \hat{a}^\dagger \hat{a} + \omega_b \hat{b}^\dagger \hat{b} + g(\hat{a}\hat{b}^\dagger+\hat{a}^\dagger \hat{b})\nonumber\\&
    + 2 E_{J_a} \cos(\varphi_\Sigma^{dc} )\sum_{k}^\infty (-1)^k\frac{\hat{\varphi}_a^{2k+1}}{(2k+1)!} + 2 E_{J_b} \eta_b \cos(\omega_d^b t)\sum_{k}^\infty (-1)^k\frac{\hat{\varphi}_b^{2k+1}}{(2k+1)!},
\end{align}
where the first line corresponds to the linear Hamiltonian and the second line describing the nonlinear part of ATS corresponding to SGS (a-mode) and buffer mode (b-mode), respectively. 

In the dispersive regime $g/(\omega_a - \omega_{b})\ll 1$, diogonalizing the linear hamiltonian provides the dressed mode of the circuit according to which we rewrite the nonlinear Hamiltonian \cite{blackbox}. To diagonalize the linear Hamiltonian, we change the mode basis, introducing the dressed modes $\{\bm{a},\bm{b}\}$ defined as $
a \equiv \varphi_a \bm{a} + \varphi_b \bm{b}, \quad b \equiv \varphi'_a \bm{a} + \varphi'_b \bm{b}.
$ Here, the coefficients are evaluated as 
$
\varphi_a \approx 1 - \mathcal{O}\left(\left(\frac{g}{\omega_a - \omega_b}\right)^2\right), \quad \varphi_b \approx \frac{g}{\omega_a - \omega_b}, 
\varphi'_b \approx 1 - \mathcal{O}\left(\left(\frac{g}{\omega_a - \omega_b}\right)^2\right), \quad \varphi'_a \approx \frac{g}{\omega_a - \omega_b},
$
corresponding to the SGS and buffer modes, respectively. 
In this new basis, the linear part of the Hamiltonian becomes diagonal and the total Hamiltonian, including the nonlinear terms and the coherent drive on the SGS, is then expressed as:
\begin{align}\label{Hamiltonian1}
    \bm{H}_{\mathrm{circuit}} = \bm{H}_{L}+\bm{H}_{\mathrm{drive}} + \bm{H}_{nL} =& \bm{\omega_a a^\dagger a }+ \bm{\omega_b b^\dagger b} + [\zeta(t) e^{i\omega_dt} \bm{a} + \zeta^*(t)e^{-i\omega_dt}\bm{a^{\dagger}}] 
    \\&+2 E_{J_a} \cos(\varphi_\Sigma^{dc} )\sum_{k}^\infty (-1)^k \varphi_{zpf}^{2k+1}\frac{(\varphi_a \bm{a}+\varphi_b \bm{b}+h.c.)^{2k+1}}{(2k+1)!} \nonumber
    \\&+ 2 E_{J_b} \eta_b \cos(\omega_d^b t)\sum_{k}^\infty (-1)^k\varphi_{zpf}^{'2k+1}\frac{(\varphi'_a \bm{a}+\varphi'_b \bm{b}+h.c.)^{2k+1}}{(2k+1)!},\nonumber
\end{align}
where we consider $\hat{\varphi}_a = \varphi_{apf}(a+a^\dagger),\hat{\varphi}_b = \varphi'_{apf}(b+b^\dagger)$ with the "\textit{zero-point-fluctuation}" coefficients $\varphi_{zpf} = \sqrt[4]{2E_c^aL_j^a},\varphi'_{zpf} = \sqrt[4]{2E_c^bL_j^b} $.
Note that the bold notation of the frequencies $\{\bm{\omega_a},\bm{\omega_b}\}$ corresponds to the dressed mode $\{\bm{a},\bm{b}\}$, respectively. The charge-line drive on the SGS is applied with the frequency $ \omega_d = n\bm{\omega}_a$ with the corresponding amplitude $\zeta(t)$ and on the buffer through the flux line with the frequency $\omega_d^b =n\bm{\omega}_a-\bm{\omega}_b $. To find the effective Hamiltonian , we apply the displacement transformation 
\begin{align}
     U = \exp(\zeta'^*(t) a- \zeta'(t)a^\dagger) \Rightarrow H_{\mathrm{dis}} = U^\dagger H U - iU^\dagger \dot{U} \rightarrow U^\dagger \dot{U} =\dot{\zeta'(t)}^* a - \dot{\zeta'(t)} a^\dagger +\frac{1}{2}(\dot{\zeta'(t)}\zeta'^*(t)-\dot{\zeta'(t)}^*\zeta'(t)),
\end{align}
where the amplitude $\zeta'(t)$ is obtained from the following relation
\begin{align}
    \dot{\zeta}'(t) = -i (\omega_a-\omega_\zeta) \zeta'(t)+ i \zeta(t).
\end{align}
In addition, in the rotating wave approximation according to $ \bm{\omega_a a^\dagger a }+ \bm{\omega_b b^\dagger b} $, the Hamiltonian \ref{Hamiltonian1} leads to
\begin{align}\label{Hamiltonian}
    \bm{H}_{\mathrm{total}} = &C_{n} (\zeta'(t) \bm{a^n} + \zeta'^*(t)\bm{a^{\dagger n}}) + g_n (\bm{a}^n\bm{b}^\dagger + \bm{a}^{\dagger n}\bm{b}),
\end{align}
where the coefficients $C_n,g_n$ are evaluated as
\begin{align}
    C_n =& 2 E_{J_a} \cos(\varphi_\Sigma^{dc})   \frac{(-1)^{n/2}\varphi_{zpf}^{n+1}\varphi_a^{n+1} }{n!} \\
    g_n =& E_{J_b} \eta_b   \frac{(-1)^{n/2}\varphi_{zpf}^{'n+1}\varphi_a^{'n}\varphi'_b }{n!}.
\end{align}

The system parameters can be chosen within the following range: \( E_c/h =  400 \) MHz, with a single junction on both ports having \( E_{J_b}/h \approx 180 \) GHz. The linear inductance in the ATS design is achieved using multiple junctions in a row, where we consider a total number of junctions \( N = 5 \) with the corresponding energy \( E'_{J_b}/h = 29 \) GHz and inductance $L_J = N/E'_J$. Since we are interested in staying within the weak drive regime, we consider the amplitude of the flux drive in the range of $\eta_b/2\pi = 0.001 - 0.04$. It is worth noting that a stronger drive amplitude would necessitate accounting for higher-order nonlinearities. Using the relation $\omega_b/2\pi =  4.3 $ GHz and considering $\omega_a/2\pi \approx 4.8 $ GHz with the coupling strength $g/2\pi\approx$   85 MHz, the effective dressed mode coefficient is evaluated as \( \varphi'_a = 0.17 \) and \( \varphi'_b = .99 \). The zero-point fluctuation is given by \( \varphi'_{\text{zpf}} = 0.6 \). The coupling strengths can be evaluated as \( g_2/2\pi = 2.7\) and \( g_4/2\pi \approx .164 \) MHz. According to Fig. 3 in the main text, the other parameters are considered as  \( (\Gamma/g_2 \approx .28, \gamma/g_2 \approx 2.5) \) and \( (\Gamma/g_4 \approx .72, \gamma/g_4 \approx 4.2 ) \),  resulting in state preparation within a total time of \( t_2 \approx 2 \mu s \) and \( t_4 \approx 9.3 \mu s\), corresponding to 2-cat and 4-cat states, respectively.


It is worth noting that we have proposed a reasonable regime of circuit parameters to provide an approximation for the total time of the release process for the examples shown in Fig. 3 in the main text. However, for a more accurate approximation, one would need to consider higher-order corrections for the rotating wave approximation and the possibility of using more advanced fabrication and design for the ATS to engineer more suitable zero-point fluctuations on both modes $\hat{a}$ and $\hat b$, as the coupling strength depends strongly on these parameters. Here we consider high $\varphi'_{zpf}$ for the buffer mode, although for the 2-cat state generation, with a smaller value, an acceptable coupling strength $g_{ab}$ can be achieved. 
To assess the feasibility of our proposal, we compare with the recent experimental paper \cite{vanselow2025} which considers an ATS as a coupler and, by applying the strongest drive, achieves a coupling strength of $g_4/2\pi \approx 0.18$ MHz. Considering higher-order approximations and more advanced ATS designs is beyond the scope of this article and can be investigated in future studies.


\section{Shortcut to Adiabaticity}

The imperfection of state generation can be attributed to photon loss into the waveguide and the buffer mode interactions during the early stages of drive pumping. To address this, accelerating the state generation process by applying a stronger drive is required. However, a strong drive may violate the adiabatic evolution condition and induce transitions to undesirable energy states. Specifically, we aim to prepare the even-parity cat states:
\begin{align}\label{C}
    \ket{C_{\alpha}^{+2}} &= \frac{\ket{\alpha(t)} + \ket{-\alpha(t)}}{\sqrt{2\big(1 + \exp(-2|\alpha(t)|^2)\big)}}, \\ 
    \ket{C_{\alpha}^{+4}} &= \frac{\ket{\alpha(t)} + \ket{-\alpha(t)} + \ket{i\alpha(t)} + \ket{-i\alpha(t)}}{2\sqrt{\big(1 + \exp(-2|\alpha(t)|^2) + 2\exp(-|\alpha(t)|^2)\cos(|\alpha(t)|^2)\big)}},
\end{align}
which can be prepared by initializing the SGS in the vacuum state \cite{SMmirrahimi2014dynamically}. 

To suppress transitions to other energy levels, we employ the so-called \textit{shortcut to adiabaticity} \cite{SMdelcampo-shortcut}. This method evaluates the counter-adiabatic Hamiltonian as
\begin{align}\label{CA}
    H_{ca} = \frac{i}{2} \bigg[\dot{\ket{\psi}}\bra{\psi} - \ket{\psi}\dot{\bra{\psi}} \bigg].
\end{align}
When the desired state is \(\ket{\psi}\), the counter-adiabatic Hamiltonian \(H_{ca}\) suppresses transitions from \(\ket{\psi}\) to other quantum states. In the following sections, we follow the method presented in \cite{SMnakamuracat} and derive analytic expressions for the counter-adiabatic terms corresponding to the generation of the 2-cat, 4-cat and pair-cat states.

\subsection{Counter adiabatic Hamiltonian for generating 2-legged cat state}\label{ca2cat}
In this section, we study the counter-adiabatic Hamiltonian to produce the 2-cat state \eqref{C}. As mentioned in the main text, the amplitude of the 2-cat state is evaluated as $\alpha(t) = e^{\frac{i3\pi}{4}}\sqrt{\frac{\Omega_d(t)\gamma}{2g^2}}$ where the corresponding coherent state in the Fock basis is
\begin{align}
    \ket{\alpha(t)} = e^{-\frac{\Omega_d(t)\gamma}{4g^2}}\sum_n e^{\frac{i3n\pi}{4}} \Big(\frac{\Omega_d(t)\gamma}{2g^2}\Big)^{n/2}\frac{1}{\sqrt{n!}}\ket{n}.
\end{align}
The time derivation of the coherent state is calculated as
\begin{align}\label{D1}
\dot{\ket{\pm\alpha(t)}} = -\frac{\dot{\Omega}_d(t)\gamma}{4g^2}\ket{\pm\alpha(t)}\pm e^{\frac{i3\pi}{4}}\frac{\dot{\Omega}_d(t)\gamma}{4g^2} \sqrt{\frac{2g^2}{\Omega_d(t)\gamma}} \hat{a}^\dagger \ket{\pm\alpha(t)}
\end{align}
and the derivation of the denominator of Eq. \eqref{C} is calculated as
\begin{align}\label{D2}
  \frac{  \frac{\dot{\Omega}_d(t)\gamma}{g^2}}{2(1+\exp(\frac{\Omega_d(t)\gamma}{g^2}))} \frac{1}{\sqrt{2(1+\exp(-\frac{\Omega_d(t)\gamma}{g^2}))}}.
\end{align}
Introducing the odd-cat basis
\begin{align}
    \ket{C^{-2}_{\alpha}} = \frac{\ket{\alpha}-\ket{-\alpha}}{\sqrt{2\big(1 - \exp(-2|\alpha(t)|^2)\big)}},
\end{align}
and utilizing the derivatives in \eqref{D1},\eqref{D2}, the time derivative of the 2-cat state is found as
\begin{align}
 \dot{ \ket{C_{\alpha}^{+2}}} = -\frac{\dot{\Omega}_d(t)\gamma}{4g^2} \tanh(\frac{\Omega_d(t)\gamma}{2g^2}) \ket{C_{\alpha}^{+2}} + e^{\frac{i3\pi}{4}} \frac{\dot{\Omega}_d(t)}{2g} \sqrt{\frac{\gamma}{\Omega_d(t)}} \sqrt{\tanh(\frac{\Omega_d(t)\gamma}{2g^2})} \hat{a}^\dagger \ket{C_{\alpha}^{-2}}.
\end{align}
Using the relations 
\begin{align}
    a  \ket{C_{\alpha}^{\pm 2}} = e^{\frac{i3\pi}{4}}\sqrt{\frac{\Omega_d(t)\gamma}{2g^2}}\sqrt{\tanh(\frac{\Omega_d(t)\gamma}{2g^2})}\ket{C_{\alpha}^{\mp 2}},\quad a^2\ket{C_{\alpha}^{\pm 2}}= e^{\frac{i3\pi}{2}}\frac{\Omega_d(t)\gamma}{2g^2}\ket{C_{\alpha}^{\pm 2}},
\end{align}
the counter adiabatic Hamiltonian \eqref{CA} is thus found as
\begin{align}\label{C3}
    H_{ca} = \frac{i}{2}  \frac{\dot{\Omega}_d(t)}{2g} \sqrt{\frac{\gamma}{\Omega_d(t)}} \sqrt{\tanh(\frac{\Omega_d(t)\gamma}{2g^2})}\Big[ e^{\frac{i3\pi}{4}}a^\dagger\ket{C_{\alpha}^{-2}} \bra{C_{\alpha}^{+2}}\Big] + h.c. \,.
\end{align}
In general, it may be hard to implement the counter adiabatic Hamiltonian, but to determine its action on the desired state $\ket{\mathrm{2cat}}_{+}$, we can exploit the relation 
\begin{align}\label{C4}
    \ket{C_{\alpha}^{+2}} \bra{C_{\alpha}^{-2}}a\ket{C_{\alpha}^{+2}} =e^{\frac{i3\pi}{4}}\sqrt{\frac{\Omega_d(t)\gamma}{2g^2}}\sqrt{\tanh(\frac{\Omega_d(t)\gamma}{2g^2})}\ket{C_{\alpha}^{+2}} = e^{\frac{-i3\pi}{4}} \sqrt{\frac{2g^2}{\Omega_d(t)\gamma}}\sqrt{\tanh(\frac{\Omega_d(t)\gamma}{2g^2})}a^2\ket{C_{\alpha}^{+2}}.
\end{align}
The first part of equation \eqref{C3} corresponds to the hermitian conjugate of Eq. \eqref{C4} and the counter adiabatic Hamiltonian is approximately obtained as
\begin{align}
    H_{ca} = \frac{i}{4} \frac{\dot{\Omega}_d(t)}{\Omega_d(t)}\tanh(\frac{\Omega_d(t)\gamma}{2g^2}) \Big[ e^{\frac{i3\pi}{2}} a^{\dagger 2} - e^{\frac{-i3\pi}{2}}a^2\Big] = \frac{1}{4} \frac{\dot{\Omega}_d(t)}{\Omega_d(t)}\tanh(\frac{\Omega_d(t)\gamma}{2g^2}) \Big[  a^{\dagger 2} + a^2\Big].
\end{align}
\subsection{Counter adiabatic Hamiltonian for generating 4-legged cat state}\label{ca4cat}
The amplitude of the coherent state corresponding to the 4-cat state is evaluated as $\alpha(t)=e^{\frac{i3\pi}{8}}\sqrt[4]{\frac{\Omega_d(t)\gamma}{2g^2}}$ with the coherent state 
\begin{align}\label{D3}
    \ket{i^k\alpha(t)} = e^{-\frac{1}{2g}\sqrt{\frac{\Omega_d(t)\gamma}{2}}}\sum_n  e^{\frac{i3n\pi}{8}} i^{kn}\Big(\frac{\Omega_d(t)\gamma}{2g^2}\Big)^{n/4}\frac{1}{\sqrt{n!}}\ket{n}, \quad k =\{0,1\}.
\end{align}
The derivative of the 4-cat state is evaluated as
\begin{align}\label{D4}
\dot{\ket{\pm i^k\alpha(t)}} = -\frac{\dot{\Omega}_d(t)}{4g}\sqrt{\frac{\gamma}{2\Omega_d(t)}}\ket{\pm i^k\alpha(t)}\pm i^k \underbrace{e^{\frac{i3\pi}{8}}\frac{\dot{\Omega}_d(t)\gamma}{8g^2} \sqrt[4]{\frac{2g^2}{\Omega_d(t)\gamma}}^3}_{B(t)} \hat{a}^\dagger \ket{\pm i^k\alpha(t)},\quad k =\{0,1\}
\end{align}
and it follows that
\begin{align}\label{4catdot}
    \dot{\ket{C_{\alpha}^{+4}}} = -\Bigg[\frac{\dot{\Omega}_d(t)}{4g}\sqrt{\frac{\gamma}{2\Omega_d(t)}} +\frac{\dot{\mathcal{N}}}{2\mathcal{N}} \Bigg] \ket{C_{\alpha}^{+4}} + B(t) \sqrt{\frac{\mathcal{N}_1}{\mathcal{N}}} a^\dagger \ket{C_{\alpha}^{+4}}_{\mathrm{basis\,1 }}.
\end{align}
where the normalization factors are \[\mathcal{N} = \Big(1+\exp(-2|\alpha(t)|^2)+2\exp(-|\alpha(t)|^2)\cos(|\alpha(t)|^2)\Big),\]
and
\[\mathcal{N}_1 = \Big(1-\exp(-2|\alpha(t)|^2)-2\exp(-|\alpha(t)|^2)\sin(|\alpha(t)|^2)\Big),\]
and the cat basis corresponds to \[\ket{C_{\alpha}^{+4}}_{\mathrm{basis\,1}} = \frac{\ket{\alpha(t)}-\ket{-\alpha(t)}+i\ket{i\alpha(t)}-i\ket{-i\alpha(t)}}{2\sqrt{\mathcal{N}_1}}.\]
It is important to mention that the 4-component cat state permits different choices of orthogonal qubit bases, and the basis of $
\{|C_{\alpha}^{+4}\rangle, |C_{\alpha}^{+4}\rangle_{\text{basis} 1}\}$ applied here differs from the basis of $
\{|C_{\alpha}^{+4}\rangle,\ket{|C_{\alpha}^{-4}} $ addressed in the main text. 


Using the relations
\begin{align}
        a  \ket{C_{\alpha}^{+4}} = e^{\frac{i3\pi}{8}}\sqrt[4]{\frac{\Omega_d(t)\gamma}{2g^2}}\sqrt{\mathcal{N}_1/\mathcal{N}}\ket{C_{\alpha}^{+4}}_{\mathrm{basis\,1}},\quad a^4\ket{C_{\alpha}^{\pm 4}}= e^{\frac{i3\pi}{2}}\frac{\Omega_d(t)\gamma}{2g^2}\ket{C_{\alpha}^{\pm 4}}
\end{align}
along with Eq. \eqref{4catdot}, the total counter adiabatic Hamiltonian \eqref{CA} is obtained as
\begin{align}
    H_{\text{ca}}^{\text{4cat}} = \frac{\dot{\Omega}_d(t)\gamma}{16g^2} \sqrt[4]{\frac{2g^2}{\Omega_d(t)\gamma}}^6 \frac{\mathcal{N}_1}{\mathcal{N}}(a^{\dagger 4} + a^4) = \frac{\dot{\Omega}_d(t)g}{4\Omega_d(t)\sqrt{2\Omega_d(t)\gamma}}\frac{\mathcal{N}_1}{\mathcal{N}}(a^{\dagger 4} + a^4).
\end{align}
Utilizing the relation
\[
\frac{\mathcal{N}_1}{\mathcal{N}} = \frac{\sinh(|\alpha(t)|^2)-\sin(|\alpha(t)|^2)}{\cosh(|\alpha(t)|^2)+\cos(|\alpha(t)|^2)},\]
the counter adiabatic terms for both 2-cat and 4-cat state can be written in a compact form
\begin{align}
    H_{\text{ca}}^{\text{ncat}}(t) = \frac{1}{4} \frac{\dot{\Omega}_d(t)}{\Omega_d(t)} \mathcal{C}_n(t)\Big[  a^{\dagger n} + a^n\Big]\Rightarrow \begin{cases}
        \mathcal{C}_n(t) = \tanh(\frac{\Omega_d(t)\gamma}{2g^2}), & n=2 \\\\\mathcal{C}_n(t) =  \frac{g}{\sqrt{2\Omega_d(t)\gamma}}\frac{\sinh(\frac{\sqrt{2\Omega_d(t)\gamma}}{2g})-\sin(\frac{\sqrt{2\Omega_d(t)\gamma}}{2g})}{\cosh(\frac{\sqrt{2\Omega_d(t)\gamma}}{2g})+\cos(\frac{\sqrt{2\Omega_d(t)\gamma}}{2g})} & n=4.
    \end{cases}
\end{align}
It is worth noting that, in our numerical simulation, we optimize two variables, $\lambda$ and $\vartheta$, to determine the Hamiltonian, $H(\lambda t) + \vartheta H_{\text{ca}}^{\text{ncat}}(\lambda t)$, maximizing the photon number in a single mode. Specifically, the parameter $\lambda$ controls the rate at which the drive is switched on and off, while the parameter $\vartheta$ alters the effect of the counter-adiabatic term on the evolution of the SGS state.

\section{Extension to Two Modes: Generation of Pair-Cat State}\label{paircat}

\begin{figure}[ht!]
\includegraphics[width=.7\textwidth]{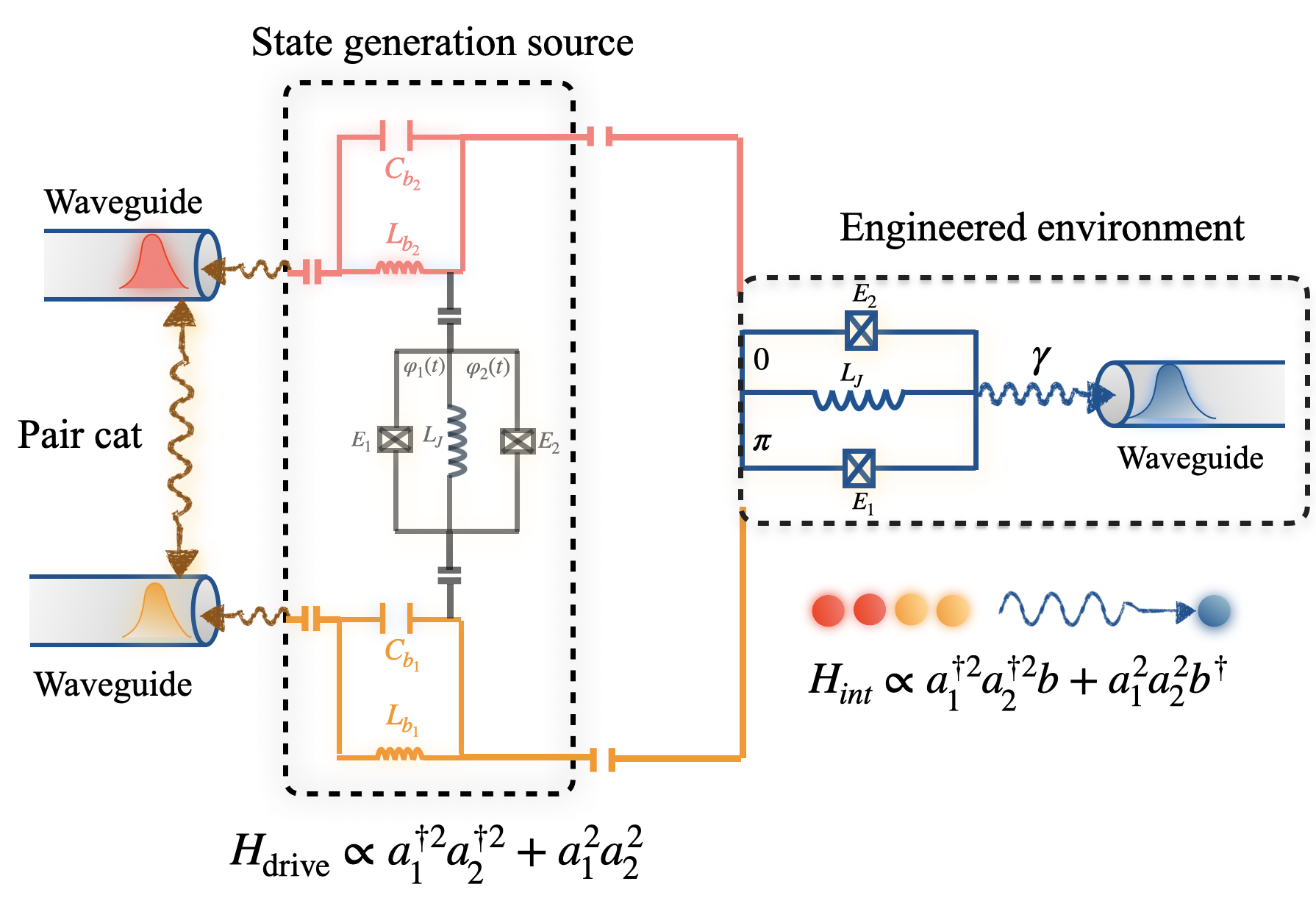}
\caption{The circuit design for generating the pair-cat state. The buffer mode, in blue interacts with two modes, \( \{a_1, a_2\} \), shown in yellow and red, respectively, through the interaction Hamiltonian \( H_{\mathrm{int}} = g \left[ a_1^{\dagger 2}a_2^{\dagger 2} b + a_1^2 a_2^2 b^\dagger \right] \). The coupler (depicted as the black ATS) is symmetrically coupled to the resonators, and by applying a proper flux drive, the joint two-photon drive \( H_{\mathrm{drive}} = \Omega(t) \left[ a_1^{\dagger 2} a_2^{\dagger 2} + a_1^2 a_2^2 \right] \) can be realized. By capturing the most populated mode in both waveguides (represented as yellow and red wavepackets on the left), the pair-cat state is successfully generated. For further details, see Sec. \ref{paircat}.
}
\label{Figpair}
\end{figure}
According to fig. \ref{Figpair}, we introduce two resonators described by operators $\{(\hat{a}_1,\hat{a}_1^\dagger),(\hat{a}_2,\hat{a}_2^\dagger)\}$ coupled to two ATS referred to a coupler and buffer mode with ladder operator $\{(\hat{c},\hat{c}^\dagger)\}$ and $\{(\hat{b},\hat{b}^\dagger)\}$, respectively. The buffer mode is the same as the two- and four-legged cat state scenario and the SGS now includes an ATS with two resonators coupled to waveguide to propagate the two-mode cat state.
Considering the frequencies $\omega_{a_1},\omega_{a_2},\omega_{b},\omega_{c}$, correspond to two resonators $a_1,a_2$, buffer mode, and coupler, respectively, and coupling strength $g_{i,j}$ between mode $i$ and $j$, the linear part of the Hmailtonian is obtained as 
\begin{align}\label{Lpair}
    H_L = &\omega_{a_1} a_1^\dagger a_1+\omega_{a_2} a_2^\dagger a_2 + \omega_b b^\dagger b + \omega_c c^\dagger c
    \nonumber\\& +g_{a_1b}(a_1b^\dagger+a_1^\dagger b)+g_{a_2b}(a_2b^\dagger+a_2^\dagger b)+ g_{a_1c}(a_1c^\dagger+a_1^\dagger c)+ g_{a_2c}(a_2c^\dagger+a_2^\dagger c),
\end{align}

thus the total Hamiltonian of the circuit in Fig. \ref{Figpair} can be written as 
\begin{align}\label{Hpair}
    H_{\mathrm{circuit}} =   H_{L}
    + 2 E_{J_c} \cos(\varphi_\Sigma^{dc} )\sum_{k}^\infty (-1)^k\frac{\hat{\varphi}_c^{2k+1}}{(2k+1)!} + 2 E_{J_b} \eta_b \cos(\omega_d^b t)\sum_{k}^\infty (-1)^k\frac{\hat{\varphi}_b^{2k+1}}{(2k+1)!},
\end{align}
Diagonalizing the linear Hamiltonian, Eq. \eqref{Lpair}, provides the dressed modes \( \{\bm{a}_1,\bm{a}_2,\bm{b},\bm{c}\} \) (bold-notation) with the relations  
\( c \equiv \varphi_{a_1} \bm{a}_1+\varphi_{a_2} \bm{a}_2+\varphi_c \bm{c}+\varphi_b \bm{b} \) and  
\( b \equiv \varphi'_{a_1} \bm{a}_1+\varphi'_{a_2} \bm{a}_2+\varphi'_c \bm{c}+\varphi'_b\bm{b} \) on the coupler and buffer mode which are utilized to effectively rewrite the nonlinear Hamiltonian in Eq. \eqref{Hpair}. Note that the coefficients $\varphi_b,\varphi'_c$ satisfy \( \varphi_b,\varphi'_c \ll 1 \), as the coupler and the buffer do not have direct connections and are assumed to be far detuned from each other.  

Similar to the scenario of 2-legged cat and 4-legged cat state generation, we apply a coherent drive through the charge-line on the coupler (ATS), given by  
\( [\zeta(t) e^{i\omega_d t} \bm{c} + \zeta^*(t)e^{-i\omega_d t}\bm{c^{\dagger}}] \),  
with the frequency \( \omega_d = 2\bm{\omega}_{a_1}+2\bm{\omega}_{a_2} \).  Additionally, we consider the flux drive on the buffer mode with frequency  
\( \omega_d^b = (2\bm{\omega}_{a_1}+2\bm{\omega}_{a_2})-\bm{\omega}_b \).  

In the rotating frame of  
\( \bm{\omega}_{a_1}\bm{a}_1^\dagger\bm{a}_1+\bm{\omega}_{a_2}\bm{a}_2^\dagger\bm{a}_2+\bm{\omega}_b\bm{b}^\dagger\bm{b}+\bm{\omega}_c\bm{c}^\dagger\bm{c} \)  
and applying the rotating wave approximation (RWA), the effective Hamiltonian of the total circuit is obtained as

\begin{align}\label{Heffpair}
    H_{\mathrm{pair-cat}} = H_{\mathrm{drive}}+H_{\mathrm{int}}= \Omega(t) [a_1^{\dagger 2}a_2^{\dagger 2} + a_1^2a_2^2] + g[ a_1^{\dagger 2}a_2^{\dagger 2} b + a_1^2a_2^2 b^\dagger]
\end{align}
where the drive and coupling coefficient are obtained as 
\begin{align}\label{pair-cpeff}
    \Omega(t) &= E_{J_c} \zeta'(t)\cos(\varphi_\Sigma^{dc}) \varphi_{zpf}^{5}\varphi_{a_{1}}^{2}\varphi_{a_2}^{2}\varphi_c\nonumber\\
    g &= E_{J_b} \eta_b \frac{\varphi_{zpf}^{'5}\varphi_{a_1}^{'2}\varphi_{a_{2}}^{'2}\varphi'_b }{2}.
\end{align}
Coupling both resonator to a waveguide with same coupling strength $\Gamma$, the total master equation is considered as
\begin{align}\label{pc}
\dot{\varrho} =-i\big[H_{\mathrm{pair-cat}},\varrho\big] +\gamma\mathcal{D}(\hat{b})\varrho + \Gamma\mathcal{D}(\hat{a}_1)\varrho+ \Gamma\mathcal{D}(\hat{a}_2)\varrho,
\end{align}
where its solution provides pair cat states.


\subsection{Counter adiabatic Hamiltonian for stabilizing pair-cat state}
We assume a symmetric interaction strength and decay rate for both modes \( \{a_1,a_2\} \), leading to the same population at each time, \textit{i.e.}, \( n_{a_1}(t) = n_{a_2}(t) \, \forall \, t \). 
Hence, the corresponding pair-cat state is proportional to the amplitude 
\[
\alpha(t) = e^{\frac{i3\pi}{8}}\sqrt[4]{\frac{\Omega_d(t)\gamma}{2g^2}}.
\]
To calculate the counter-adiabatic Hamiltonian in Eq. \eqref{CA} for the pair-cat state production, one needs to evaluate the time derivative of the state:
\begin{align}
    \ket{\psi}_+ = \frac{\ket{\alpha(t),\alpha(t)}+\ket{i\alpha(t),i\alpha(t)}}{\mathcal{N}_+(t)}
\end{align}
where the pair-coherent state is given by
\begin{align}
\ket{\alpha(t),\alpha(t)} = \frac{1}{\mathcal{N}(t)}\sum_{n=0}^\infty\frac{\alpha^{2n}(t)}{n!} \ket{nn} \Rightarrow \mathcal{N}(t) = \sqrt{I_0(2|\alpha(t)|^2)}.
\end{align}
Here, \( I_0(z) = \sum_{k=0}^\infty\frac{(z/2)^{2k}}{k!^2} \) is the modified Bessel function of the first kind, and the normalization factor of the pair-cat state is 
\[
\mathcal{N}_\pm(t) = \sqrt{2\left(1\pm \frac{J_0(2|\alpha(t)|^2)}{I_0(2|\alpha(t)|^2)}\right)},
\]
where \( J_0(z) =\sum_{k=0}^\infty(-1)^k\frac{(z/2)^{2k}}{k!^2} \) corresponds to the Bessel function of the first kind.

The time derivative of the pair-coherent states is evaluated as
\begin{align}
    \dot{\ket{\alpha(t),\alpha(t)}} &= -\frac{\dot{\mathcal{N}}(t)}{\mathcal{N}(t)}\ket{\alpha,\alpha} + 2 \alpha(t) \dot{\alpha}(t) a_{1}^\dagger a_{2}^\dagger \ket{\alpha,\alpha},\nonumber\\
    \dot{\ket{i\alpha(t),i\alpha(t)}} &= -\frac{\dot{\mathcal{N}}(t)}{\mathcal{N}(t)}\ket{i\alpha, i\alpha} - 2 \alpha(t) \dot{\alpha}(t) a_{1}^\dagger a_{2}^\dagger \ket{i\alpha, i\alpha}.
\end{align}
This provides the time derivative of the pair-cat state as follows:
\begin{align}
    \dot{\ket{\psi}_+} = -\left(\frac{\dot{\mathcal{N}}(t)}{\mathcal{N}(t)}+\frac{\dot{\mathcal{N}}_+(t)}{\mathcal{N}_+(t)}\right)\ket{\psi}_+ + 2 \alpha(t) \dot{\alpha}(t) a_{1}^\dagger a_{2}^\dagger \frac{\mathcal{N}_-(t)}{\mathcal{N}_+(t)} \ket{\psi}_-.
\end{align}
Considering Eq. \eqref{CA}, the counter-adiabatic Hamiltonian simplifies to 
\begin{align}
    H_{\mathrm{ca}}^{\mathrm{pair}} = \frac{i}{2} \left(2 \alpha(t) \dot{\alpha}(t) a_{1}^\dagger a_{2}^\dagger \frac{\mathcal{N}_-(t)}{\mathcal{N}_+(t)} \ket{\psi}_- \bra{\psi}_+ - \text{h.c.} \right),
\end{align}
where the effect of the second term of the Hamiltonian on the state \( \ket{\psi}_+ \) can be expressed as
\begin{align}
    2 \alpha^*(t) \dot{\alpha}^*(t) \frac{\mathcal{N}_-(t)}{\mathcal{N}_+(t)}\ket{\psi}_+ \bra{\psi}_- ab \ket{\psi}_+ &\equiv 2 \alpha^*(t) \dot{\alpha}^*(t) \frac{\mathcal{N}_-(t)^2}{\mathcal{N}_+(t)^2} \alpha^2(t) \ket{\psi}_+ \nonumber\\
    &\equiv 2 \frac{\alpha^*(t) \dot{\alpha}^*(t)}{\alpha^2(t)} \frac{\mathcal{N}_-(t)^2}{\mathcal{N}_+(t)^2} a_{1}^2a_{2}^2 \ket{\psi}_+ = \mathcal{G}(t)a_{1}^2a_{2}^2 \ket{\psi}_+.
\end{align}
Hence, the counter-adiabatic Hamiltonian can be approximated as
\begin{align}
    H_{\mathrm{ca}}^{\mathrm{pair}} = \frac{i}{2} \left(\mathcal{G}^*(t) a_{1}^{\dagger 2}a_{2}^{\dagger 2} - \mathcal{G}(t) a_{1}^2a_{2}^2\right) \Rightarrow H_{\mathrm{ca}}^{\mathrm{pair}} = \frac{1}{4} \frac{\dot{\Omega}(t)}{\Omega(t)} \frac{\mathcal{N}_-(t)^2}{\mathcal{N}_+(t)^2} (a_{1}^{\dagger 2}a_{2}^{\dagger 2} + a_{1}^2a_{2}^2).
\end{align}
We add this term to the Hamiltonian in Eq. \eqref{pc} to simulate the production of the pair-cat state.

\section{Generating single-mode propagating grid state}
\begin{figure}[ht!]
\includegraphics[width=.7\textwidth]{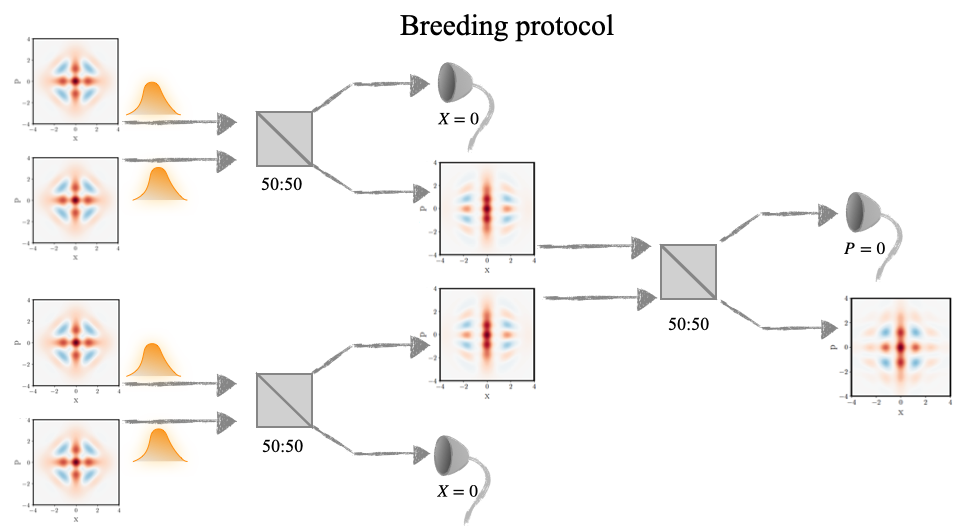}
\caption{The breeding protocol is implemented to generate the propagating grid state. 
The first iteration involves projecting the output port of the beamsplitter on an eigenstate of the $X$ quadrature operator, 
followed by a second iteration where a projection is performed on an eigenstate of the $P$ quadrature operator.
}
\label{Fig3}
\end{figure}

If we consider a bosonic mode with quadratures $\hat{q} = \frac{1}{\sqrt{2}} (a^\dagger + a)$ and $\hat{p} = \frac{i}{\sqrt{2}} (a^\dagger - a)$, 
the grid states in this bosonic mode correspond to the $+1$ eigenstates of the commuting operators $S_q = e^{iu\hat{q}}$ and $S_p = e^{iv\hat{p}}$, 
where $[S_q, S_p] = 0$.
The condition $uv \, \mathrm{mod} \, 2\pi = 0$ ensures the commutativity of $S_q$ and $S_p$, and without loss of generality, 
we take $u = v = \sqrt{2\pi}$.
The ideal eigenstates of these operators have infinite energy, which is physically unattainable. Therefore, a realistic grid state is introduced 
with a finite photon number, implemented via a Gaussian envelope in the Fock basis \cite{SMPhysRevA.95.012305,SMterhalGKP} as follows,
\begin{align}\label{finitgkp}
    \ket{\psi} \propto \sum_{m=-\infty}^{\infty}e^{-\pi\Delta^2 m^2}\hat{D}(m\sqrt{\pi})\hat{S}(\Delta)\ket{0},
\end{align}
where $\hat{S}(\Delta) = \exp(\Delta (a^2-a^{\dagger 2})) $ and $\hat{D}(\alpha) = \exp(\alpha a^\dagger - \alpha^* a)$ with a distance $\sqrt{2\pi}$ between it picks in the phase space. As mentioned in the main text, one parameter used to quantify the quality of the grid state is the effective squeezing \cite{terhalGKP}, defined as:
\begin{align}
    \Delta_s = \frac{1}{\sqrt{\pi}} \sqrt{\ln\Big(\mathrm{Tr}[\hat{D}(\sqrt{\pi})\varrho]^{-2}\Big)},
\end{align}
where $\Delta_s$ can be reported in decibels (dB) as follows:
\begin{align}
    \Delta = -10 \log_{10}(\Delta_s^2/\Delta^2_{0}),
\end{align}
where $\Delta_0{^2} = 0.5$ represents the quadrature variance of the vacuum states.
\begin{figure}[ht!]
\includegraphics[width=.9\textwidth]{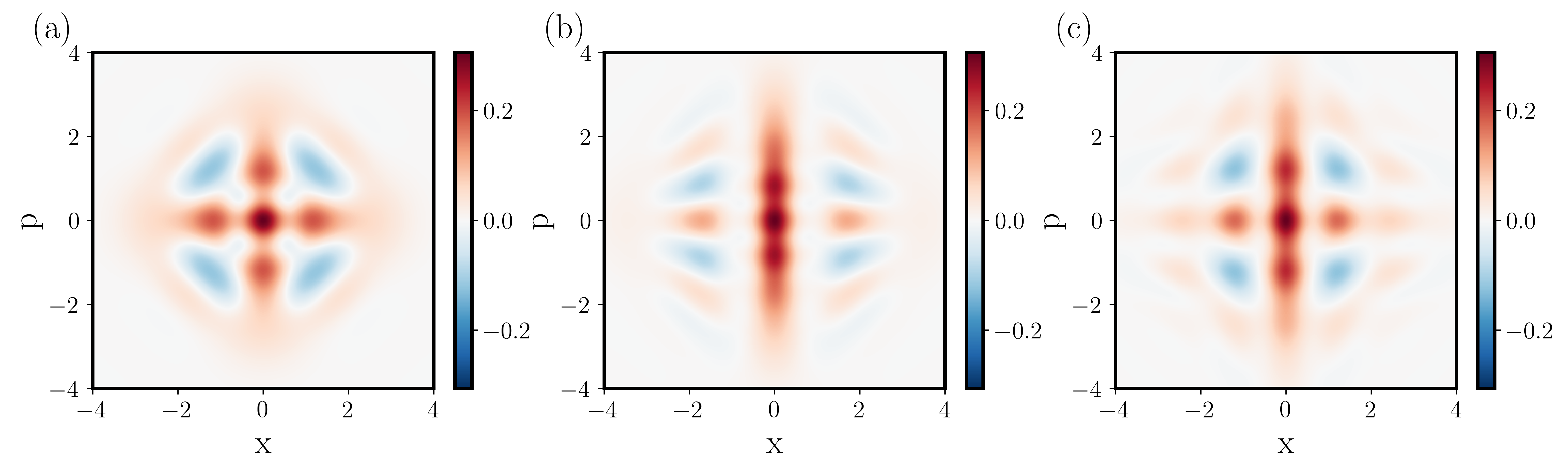}
\caption{The output of the first (b) and second (c) iterations of the breeding protocol applied to the 4-cat state on panel (a). As mentioned in the main text, panel (a) represents a 4-cat state 
with $|\alpha|^2 = 2.1$ and a fidelity of 95\%. Panels (b) and (c) show the conditional state resulting from a measurement 
of $x = 0$ ($p = 0$) on the other output port of the beam splitter, respectively. The effective 
squeezing in panel (c) is $\Delta_p = 1.1$ dB and $\Delta_x = 3.6$ dB.}
\label{Fig4}
\end{figure}
To prepare such a grid state, one can apply the iterative breeding protocol on squeezed cat states, 4cat, and on binomial states \cite{SMterhalGKP,SMPhysRevA.108.012603}. 
A single iteration of the breeding protocol involves a 50:50 beamsplitter, followed by a projection measurement, assuming 
an ideal homodyne measurement, on $x $ or $p$. In this work, we follow the protocol outlined in \cite{SMPhysRevA.108.012603}, 
with the first projection on the position operator $x=0$, followed by a subsequent projection on the momentum operator $p = 0$.
It should be noted that more complex breeding protocols have been studied and could be utilized here \cite{PhysRevA.110.012436}. 
However, analyzing the characteristics of these protocols is beyond the scope of this paper. In Fig. ~\ref{Fig4}, three Wigner functions of the grid state have been shown. Panel (a) shows the state occupying the most populated mode generated by the SGS and panel (b) and (c) are the conditional quantum state on projective measurement in $x=0$ and $p=0$, respectively.